\documentclass[%
 reprint,
 superscriptaddress,
 nofootinbib,
 amsmath,amssymb,
 aps,
]{revtex4-2}

\usepackage{graphicx}
\usepackage{dcolumn}
\usepackage{bm}
\usepackage{xcolor}
\usepackage{amsmath}
\usepackage{appendix}
\usepackage{hyperref}

\newcommand{\tresid}[0]{\delta\mathbf{t}}
\newcommand{\fmat}[0]{\mathbf{F}}
\newcommand{\tmat}[0]{\mathbf{T}}
\newcommand{\avec}[0]{\mathbf{a}}
\newcommand{\rvec}[0]{\mathbf{r}}
\newcommand{\mmat}[0]{\mathbf{M}}
\newcommand{\nmat}[0]{\mathbf{N}}
\newcommand{\bvec}[0]{\mathbf{b}}
\newcommand{\bmat}[0]{\mathbf{B}}

\newcommand{\cmat}[0]{\mathbf{C}}
\newcommand{\dmat}[0]{\mathbf{D}}
\newcommand{\epsvec}[0]{\boldsymbol{\epsilon}}
\newcommand{\phimat}[0]{\boldsymbol{\varphi}}
\newcommand{\Phimat}[0]{\boldsymbol{\Phi}}
\newcommand{\etavec}[0]{\boldsymbol{\eta}}
\newcommand{\rhovec}[0]{\boldsymbol{\rho}}

\usepackage{macros_ads}

\begin{document}

\title{Posterior predictive checking for gravitational-wave detection with pulsar timing arrays:
II. Posterior predictive distributions and pseudo Bayes factors}

\author{Patrick M. Meyers}
\email{pmeyers@caltech.edu}
\affiliation{Department of Physics, California Institute of Technology, Pasadena, California 91125, USA}

\author{Katerina Chatziioannou}
\email{kchatziioannou@caltech.edu}
\affiliation{LIGO  Laboratory,  California  Institute  of  Technology,  Pasadena,  California  91125,  USA}
\affiliation{Department of Physics, California Institute of Technology, Pasadena, California 91125, USA}

\author{Michele Vallisneri}
\email{Michele.Vallisneri@jpl.nasa.gov}
\affiliation{Jet Propulsion Laboratory, California Institute of Technology, Pasadena CA 91109, USA}
\affiliation{Department of Physics, California Institute of Technology, Pasadena, California 91125, USA}

\author{Alvin J. K. Chua}
\email{alvincjk@nus.edu.sg}
\affiliation{Department of Physics, National University of Singapore, Singapore 117551}
\affiliation{Department of Mathematics, National University of Singapore, Singapore 119076}

\date{\today}

\begin{abstract}
The detection of nanoHertz gravitational waves through pulsar timing arrays hinges on identifying a common stochastic process affecting all pulsars in a correlated way across the sky. In the presence of other deterministic and stochastic processes affecting the time-of-arrival of pulses, a detection claim must be accompanied by a detailed assessment of the various physical or phenomenological models used to describe the data. In this study, we propose \textit{posterior predictive checks} as a model-checking tool that relies on the predictive performance of the models with regards to new data. We derive and study predictive checks based on different components of the models, namely the Fourier coefficients of the stochastic process, the correlation pattern, and the timing residuals. We assess the ability of our checks to identify model misspecification in simulated datasets. We find that they can accurately flag a stochastic process spectral shape that deviates from the common power-law model as well as a stochastic process that does not display the expected angular correlation pattern.
Posterior predictive likelihoods derived under different assumptions about the correlation pattern can further be used to establish detection significance.
In the era of nanoHertz gravitational wave detection from different pulsar-timing datasets, such tests represent an essential tool in assessing data consistency and supporting astrophysical inference.
\end{abstract}

\maketitle

\section{\label{sec:intro}Introduction}

Building on millisecond-pulsar observations spanning decades, four international pulsar-timing-array (PTA) collaborations have recently reported varying levels of evidence for a low-frequency gravitational-wave (GW) background~\cite{NANOGrav:2023gor,EPTA:2023fyk,Reardon:2023gzh,Xu:2023wog},
which is broadly expected from the binaries of supermassive black holes at the centers of galaxies~\cite{NANOGrav:2023hfp,EPTA:2023xxk,EPTA:2023gyr,NANOGrav:2023tcn,NANOGrav:2023pdq}, but may also have been generated by ``new physics''~\cite{NANOGrav:2023hvm,EPTA:2023xxk}.
The PTAs are now collaborating to compare their estimates of the amplitude, shape, and significance of the background~\cite{InternationalPulsarTimingArray:2023mzf}.

All PTAs use similar data models, which typically include
a deterministic timing model characterizing the motion of each pulsar~\cite{Edwards:2006zg,Luo:2020ksx}, stochastic noise that affects each pulsar individually (dispersion measure fluctuations~\cite{Goncharov:2020krd,Jones:2016fkk} and intrinsic pulsar red noise~\cite{Groth1975,Cordes1980,Shannon:2010bv}), a GW background common to all pulsars, as well as measurement noise. The intrinsic pulsar noise and the GW background are modeled phenomenologically as finite Gaussian processes with Fourier bases functions and power-law priors \cite{Phinney:2001di,2013PhRvD..87j4021L,vanHaasteren:2014qva,vanHaasteren:2008yh}, although more complex models have been proposed \cite{NANOGrav:2021yqt,Lam:2017duo,Goncharov:2020krd}.
Given that GW searches rely crucially on these phenomenological models, it is important to develop methods to identify and assess model misspecification.

The most common model-checking approach consists of  modifying parts of a model and then comparing the ratio of the marginal likelihoods (i.e., the Bayes factor) between the original and modified models.
However, there are two problems with adopting the Bayes factor for this task. The first is a problem of principle: in addition to a Bayes factor, model comparison requires prior odds. However, it seems very hard to assign priors to hypotheses about the very existence of the GW background and its spectral shape, or to the  unphysical null models used to establish detection significance.
Furthermore, no set of models exhausts the space of relevant hypotheses, which should include alternatives that embody known and unknown systematics; indeed, a faithful model may be impossible to specify formally~\cite{bernardo2009bayesian}.
The second problem is one of interpretation: even taking model comparison at face value, it remains unclear what confidence a Bayes factor actually conveys beyond arbitrary mappings~\cite{jeffreys1998theory,kassraftery1995} of Bayes factors to degree-of-belief descriptors (``strong'', ``decisive'', etc.).

Such issues aside, the central idea of model checking through Bayesian model comparison has been thoroughly explored and employed in PTA analyses.
In the parlance of hierarchical inference~\cite{Loredo2004,Mandel2019}, the description of the pulsar noise and GW background by the Gaussian-process likelihood and decomposition onto sinusoids is the \textit{model}, while the (complex) amplitude of each sinusoid is a \textit{parameter} of the model. The assumption that the amplitudes follow a power-law is the \textit{hierarchical model} (or \emph{hypermodel}), and the amplitude and spectral index of the power-law are \textit{hyperparameters}. In this context, the most straightforward check involves changing elements of the (hyper)model~\cite{Sampson:2015ada,Hazboun:2020kzd,Goncharov:2022ktc,Zic:2022sxd}. For example, Ref.~\cite{Sampson:2015ada} replaced the power-law with a truncated power-law and Ref.~\cite{Zic:2022sxd} explored the impact of the hyperparameter priors on the marginal likelihoods.  However, since the model and the hypermodel for the stochastic processes are mainly phenomenological and unlikely to be perfectly representing reality, model comparisons between these extensions do not have a clear interpretation.
 
We propose a complementary approach to assessing model misspecification that hinges on the predictive power of our analysis with regards to new data. In a companion paper~\cite{ppc1} we explore predictive tests in the context of null-hypothesis testing with the \textit{optimal statistic}~\cite{Anholm:2008wy,2015PhRvD..91d4048C,Vigeland:2018ipb,Hazboun:2023tiq}; by contrast, this study focuses on Bayesian inference. The main idea behind predictive tests is to use inference based on current data to predict further data. Comparing the prediction with future or current data then allows us to probe different elements of the analysis. Compared to tests based on perturbing a model and comparing the marginalized likelihoods, predictive tests focus naturally on specific elements of the model or the hypermodel. For example, predictive checks of the GW spectrum allow us to directly assess whether specific frequency components have been over- or under-estimated.  
In the context of GW analyses, such tests are a common step of estimating the populations of binary black hole and binary neutron star systems~\cite{Fishbach:2019ckx,Callister:2022qwb,LIGOScientific:2018jsj,LIGOScientific:2020kqk}. Similar posterior predictive tests have been used on individual pulsars~\cite{Wang:2019xfx}; our study here applies these tools to full PTA data analysis.

Following the discussion of~\cite{Fishbach:2019ckx} we identify three types of predictive tests, each targeting a different element of the analysis. 
\begin{itemize}
\item The first and least explored test relies on the hyperparameters (e.g., the GW background amplitude and spectral slope). For example, the inferred hyperparameters from one PTA dataset (say, NANOGrav), can be used to predict data and inference products for another dataset (say, PPTA), which can then be compared with the actual data and products. 
We leave the detailed exploration of these to future work.
\item The second test is based on the model parameters and specifically the Gaussian-process coefficients (i.e., the Fourier-component amplitudes). We consider these coefficients under two probability distributions. \textit{Predicted} coefficients are conditioned on the hypermodel and the posterior for the hyperparameters given the observed data:  for instance, for a power-law background model, they would span the range of GW signals expected given the amplitude and spectral slope inferred from the data.
\textit{Inferred} coefficients are conditioned on both the posterior of the hyperparameters \emph{and} the data:
 for a power-law background model, they would span the range of GW-induced residuals that are compatible with the data under the power-law assumption.
 By comparing predicted and inferred coefficients, we are considering whether the Fourier amplitudes actually follow a powerlaw with an assumed correlation pattern. 
\item The third test examines the pulsar-timing residual data directly through leave-one-out cross-validation on the population of pulsars. That is, we use $N_{\rm p}-1$ pulsars to calculate the (posterior predictive) likelihood of the data observed for the $N_{\rm p}^{\mathrm{th}}$ pulsar.
We assess the likelihoods in the context of model criticism, (which pulsars are not predicted well by the model fit to the other pulsars?), and model comparison (which model, fit to $N_{\rm p}-1$ pulsars, does best at predicting the residuals of the left-out pulsar?). We further propose that a summary statistic built from the posterior predictive likelihoods can be used to establish detection significance, by comparing its observed value to a null distribution obtained from simulated datasets with no GW background.
\end{itemize}
 
We assess tests on the Gaussian process coefficients using simulated datasets that represent different levels of model misspecification.
Simulations are based on the times-of-arrival (TOAs) and noise parameters of the NANOGrav 12.5-yr dataset~\cite{NANOgrav:12p5yrData} to create synthetic residuals and include a GW signal. We consider (i) a dataset that obeys our assumptions of a GW background with a power-law spectral shape and Hellings--Downs correlations; (ii) a dataset that breaks the power-law assumption, instead having a truncated power-law spectral shape; and (iii) a dataset that breaks the correlation-pattern assumption by adding monopolar correlations. Comparing inferred and predicted coefficients allows us to identify model misspecification for both (ii) and (iii).

Switching to predictive tests with the timing residuals, we introduce a ``pseudo Bayes factor"~\cite{geiser_eddy_pbf}, defined as the ratio of the posterior predictive likelihoods of the observed data in a pulsar given all other pulsars under a model that includes Hellings--Downs correlations and a model that assumes no spatial correlations. 
We compute the pseudo Bayes factor for simulated datasets that contain a GW background and for ``null" simulations with no signal. We show that, similar to the standard drop-out factor~\cite{NANOGrav:2020bcs}, the pseudo Bayes factor is an indicator of Hellings--Downs correlations in most pulsars.
However, even in the presence of a signal some pulsars show preference against Hellings--Downs. The latter seems to be an expected feature of PTA datasets. Finally, we compare the \emph{total} pseudo Bayes factor, i.e., the product over all pulsars, between the datasets with and without a GW and show that it can be used as a detection statistic. 

The rest of this article is organized as follows. In Sec.~\ref{sec:pta_likelihood} we summarize PTA analyses. In Sec.~\ref{sec:pch} we comment on posterior predictive checks using hyperparameters. In Secs.~\ref{sec:results-parameters} and~\ref{sec:results-parameters} we propose and test posterior predictive checks for model parameters and timing data respectively. In Sec.~\ref{sec:discussion_and_conclusion} we conclude.

\section{\label{sec:pta_likelihood}Pulsar timing array analysis}

We begin with an overview of PTA analysis with an emphasis on the modeling choices we test in the subsequent sections. For a more detailed discussion on PTA physics and analyses, see Refs.~\cite{Taylor:2021yjx,vanHaasteren:2012hj,NANOGrav:2015aud}. 

\subsection{PTA model and likelihood}
\label{sec:model}

The arrival times of radio pulses are influenced by both deterministic and stochastic processes. Deterministic effects include the apparent and proper motion of the pulsar, as well as its orbit in a binary. A first analysis step fits a \textit{timing model} that describes the deterministic effects and subtracts it from the arrival times to obtain the \textit{timing residuals} $\tresid$~\cite{Edwards:2006zg,Luo:2020ksx}. Recovery of the best-fit timing model is influenced by stochastic processes such as \textit{spin noise}~\cite{Groth1975,Cordes1980,Shannon:2010bv} -- 
stochastic fluctuations of the pulsar rotation frequency intrinsic to each individual pulsar -- and GWs, which induce a correlated stochastic signal common to all pulsars.
For example, red noise affects (among others) the estimate of the pulsar rotation period and its derivative.

Assuming that the effect of stochastic processes on the timing solution is small, most PTA analyses are based on the timing residuals $\tresid$, which we use here to denote timing residuals for all pulsars concatenated into a single vector. 
Stochastic processes are modeled in terms of their frequency content, expressed through a matrix $\fmat$ that contains sines and cosines of different frequencies and a vector of amplitudes $\avec$ associated with each frequency~\cite{2013PhRvD..87j4021L}.\footnote{Time-domain approaches have also been considered~\cite{vanHaasteren:2008yh}.} 
Additionally, the presence of red noise in the original arrival times will have shifted the best-fit timing solution from its ``true'' value. We correct for this effect within a linear approximation, with a known design matrix $\mmat$ of partial derivatives mapping small changes in timing model parameters $\epsvec$ onto changes in $\tresid$.
Defining
\begin{align} 
\tmat & = \begin{bmatrix}\mmat & \fmat\end{bmatrix}\,,\label{eq:t_matrix}\\
\bvec &= \begin{bmatrix}\epsvec \\ \avec \end{bmatrix}\,,\label{eq:gp_coefficients}
\end{align}
the full model residuals are
\begin{align}
\rvec &= \tresid - \tmat\bvec\,,\label{eq:residuals}
\end{align}
and under the assumption of Gaussian measurement noise the likelihood is the Gaussian distribution
\begin{align}
p(\tresid |\bvec) = \frac{\exp\left(-\frac{1}{2}\rvec ^T\nmat^{-1}\rvec\right)}{\sqrt{\det\left(2\pi\nmat\right)}}\,.
\label{eq:pre_fit_resids_likelihood_gp_coefficients}
\end{align}
For ``narrowband'' timing campaigns, $\nmat$ is a block-diagonal noise matrix in which the dense blocks arise due to pulse profile ``jitter'' noise that is correlated across arrival times taken at different radio frequency channels during the same observation~\cite{NANOGrav:2015qfw}. If TOAs across the measurement band are condensed into single TOAs, $\nmat$ is diagonal. In what follows, we assume $\nmat$ is characterized accurately and we do not consider relevant mismodeling. 

At this stage, the model parameters are the sine and cosine spectral amplitudes $\avec$ and the timing model corrections, $\epsvec$, though we are primarily interested in the former. In order to separate the intrinsic pulsar noise and the common GW, we place a Gaussian hyperprior on $\bvec$ in terms of the hyperparameters $\bm \Lambda$

\begin{align}
    p(\bvec | \bm \Lambda) &= \frac{\exp\left(-\frac{1}{2}\bvec^T \bmat^{-1}\bvec\right)}{\sqrt{\det(2\pi\bmat)}}\label{eq:model_parameters_hyper_prior}\,,\\
    \textrm{with }\bmat &= \begin{bmatrix}
    \infty & 0 \\
    0 & \phimat(\bm \Lambda)
    \end{bmatrix}\,.\label{eq:big_b_matrix_definition}
\end{align}
The top-corner entries of $\bmat$ express an improper prior of infinite variance on the timing-model corrections $\epsvec$. The matrix $\phimat(\bm \Lambda)$ includes the correlation of different elements of $\bvec$ via power spectra  $\etavec(\bm \Lambda)$ and $\rhovec(\bm \Lambda)$ that encode the intrinsic pulsar noise and the GW signal respectively. Furthermore, GWs induce correlations in the same frequency bin for different pulsars based on their angular separation as prescribed by the Hellings--Downs curve. Overall for each of the sine and cosine coefficient in $\avec$,
\begin{align}
    \phimat(\bm \Lambda)_{(ai, bj)} = \Gamma_{ab}\rho_i^2(\bm \Lambda)\delta_{ij} + \eta_{ai}^2(\bm \Lambda) \delta_{ab}\delta_{ij}\,,\label{eq:phi_matrix_definition}
\end{align}
where $a$ and $b$ label pulsars, and $i$ and $j$ label frequencies. The GW power spectrum at a given frequency is captured by $\rho_i(\bm \Lambda)$, the Hellings--Downs curve by $\Gamma_{ab}$, and the power spectrum of the intrinsic pulsar noise associated with each individual pulsar at each individual frequency by $\eta_{ai}(\bm \Lambda)$. A stronger assumption is that both $\eta_{ai}(\bm \Lambda)$ and $\rho_i(\bm \Lambda)$ follow a power-law
\begin{align}
    \rho_{i}^2(\bm \Lambda) &= \frac{A_{\textrm{gw}}^2}{12\pi^2}\left(\frac{f_i}{f_{\textrm{y}}}\right)^{-\gamma_{\textrm{gw}}}\frac{f_{\textrm{y}}^{-3}}{T}\,,\label{eq:gw_power_law_definition}\\
    \eta_{ai}^2(\bm \Lambda) &= \frac{A_{a,\textrm{int}}^2}{12\pi^2}\left(\frac{f_i}{f_{\textrm{y}}}\right)^{-\gamma_{a, \textrm{int}}}\frac{f_{\textrm{y}}^{-3}}{T}\,,\label{eq:red_noise_definition}
\end{align}
where $A_{\textrm{gw}}$ is the amplitude of the GW background at $f_\textrm{y}$, $f_i= i / T$ is the frequency of the $i^{\textrm{th}}$ bin, $f_{\textrm{y}} = (1\,\textrm{y})^{-1}$, and $T$ is the dataset duration. Throughout, we use $i \in [1\mbox{--}10]$ for the GW background ($f = 2.5\mbox{--}24.6\;\mathrm{nHz}$) and $i \in [1\mbox{--}30]$ for the intrinsic red noise\footnote{We use 10 frequencies for the GW background as opposed to the 5 frequencies used in~\cite{NANOGrav:2020bcs} because we have injected a signal that is stronger than the common process observed in that analysis.}.

Under the power-law assumption, the model \textit{hyperparameters} $\bm \Lambda$ are the GW amplitude $A_{\textrm{gw}}$ and the spectral index $\gamma_{\textrm{gw}}$, and an intrinsic pulsar noise amplitude $A_{a,\textrm{int}}$ and spectral index $\gamma_{a, \textrm{int}}$ for each of the $N_{\textrm{p}}$ pulsars. The posterior on these hyperparameters is obtained by marginalizing over the model parameters $\bvec$,
\begin{align}
    p(\bm \Lambda | \tresid) &= \int d\bvec\,p(\tresid | \bvec)p(\bvec | \bm \Lambda)p(\bm \Lambda)\nonumber\\
    &= \frac{p(\bm \Lambda)}{\sqrt{\det(2\pi \cmat)}}\exp\left(-\frac{1}{2} \tresid^T \cmat^{-1}\tresid\right)\,,\label{eq:hyper_parameter_posterior}
\end{align}
where the new covariance matrix is $\cmat\equiv\left(\nmat + \tmat \bmat \tmat^T\right)$, and $p(\bm \Lambda)$ is the prior on the hyperparameters. Alternatively, the first two terms in the integrand of Eq.~\eqref{eq:hyper_parameter_posterior} can be written as a posterior, $p(\bvec | \tresid,\bm\Lambda)$, which is normal with mean and covariance given respectively by
\begin{align}
\label{eq:bvec_max_likelihood}
    \mathbf{\widehat{b}} &= \boldsymbol\Sigma\tmat^T\nmat^{-1}\tresid\,, \\
    \label{eq:bvec_covariance_matrix}
    \boldsymbol\Sigma &= \left(\tmat^{T}\nmat^{-1}\tmat + \bmat^{-1}\right)^{-1}\,.
\end{align}

Given the large dimensionality ($2 N_{\textrm{p}} + 2$ hyperparameters for a typical analysis), most GW analyses estimate the marginalized posterior on the hyperparameters $\bm \Lambda$ through stochastic sampling, resulting in $N_s$ samples $\left\{\bm\Lambda^{s}\right\}_{s=1}^{N_s}$ drawn from their posterior,
\begin{align}
\label{eq:hyper_parameter_posterior_samples}
    \bm\Lambda^{s} &\sim p(\bm \Lambda | \tresid)\,.
\end{align}
In Section \ref{sec:pch}, we propose methods to assess how well the models and assumptions of this section fit the data based on having obtained $\bm\Lambda^{(s)}$.

\subsection{Simulated datasets}
\label{sec:datasets}

We experiment with our proposed methods by analyzing simulated datasets. We consider a total of four datasets, each spanning 12.9 years of data over 45 pulsars, and produce one realization for each of those datasets.
\begin{itemize}
    \item \textsc{HellingsDowns-PowerLaw}: Constructed in accordance with the assumptions described in Sec.~\ref{sec:model}, this dataset contains a GW signal described by a power-law with $\log_{10} A_{\textrm{gw}}=-14$ and $\gamma_{\textrm{gw}}=13/3$, see Eq.~\eqref{eq:gw_power_law_definition}. The Hellings--Downs correlations are detectable with an optimal-statistic signal-to-noise ratio (SNR) of $5.5$. 
    \item \textsc{HellingsDowns-Turnover}: Constructed to test the power-law assumption, this dataset contains a GW signal described by the broken power-law
    \begin{equation}
        \rho^2(f) = \frac{A_{\textrm{gw}}^2}{12\pi^2}\left(\frac{f}{f_{\textrm{yr}}}\right)^{\!\!-\gamma_{\textrm{gw}}}\left[1 + \left(\frac{f_{\textrm{b}}}{f}\right)^{\!\!\kappa}\right]^{-1}\frac{f_{\textrm{y}}^{-3}}{T},
    \end{equation}
    with $\gamma_{\textrm{gw}}=13/3$,  $\log_{10}A_{\textrm{gw}}=-13.5$, $f_{\textrm{b}}=7.9$\,nHz, and $\kappa=26/3$.\footnote{This value is chosen for illustrative purposes, as it produces a noticeable turnover at low frequencies. It does not correspond to a specific astrophysical scenario.} The optimal statistic SNR is $4.4$.
    \item \textsc{HellingsDownsMonopole-PowerLaw}: The third dataset focuses on spatial correlations and includes a power-law GW signal with $\log_{10} A_{\textrm{gw}}=-14$ and $\gamma_{\textrm{gw}}=13/3$ as well as a stochastic process with $\log_{10} A_{\textrm{m}}=-14.3$ and $\gamma_{\textrm{m}}=13/3$ that induces monopolar correlations across the pulsars ($\Gamma_{ab}=1$).
    The optimal-statistic SNR is $6$.\footnote{This SNR is calculated assuming only Hellings--Downs correlations.}
    \item \textsc{NoGravitationalWave}: Finally, we consider a dataset without any common process between the pulsars, setting $ A_{\textrm{gw}}=0$.
\end{itemize}
Hyperparameters for the intrinsic pulsar noise are chosen from the posteriors of the NANOGrav 12.5-yr dataset~\cite{NANOGrav:2020gpb,NANOgrav:12p5yrData}. 
We simulate data by first drawing from the posterior distribution on the intrinsic pulsar noises $A_{a,\textrm{int}}^{\textrm{sim}},\gamma_{a,\textrm{int}}^{\textrm{sim}} \sim p(A_{a,\textrm{int}},\gamma_{a,\textrm{int}} | \tresid^{\textrm{NG12.5}})$. The GW parameters are specified independently and listed above, thus completing the list of simulated hyperparameters $\bm \Lambda^{\textrm{sim}}$. We then draw Gaussian process coefficients as $\avec^{\textrm{sim}}\sim p(\avec | \bm\Lambda^{\textrm{sim}})$ and set the timing parameters $\bm \epsilon^{\textrm{sim}}= 0$. Finally, we draw simulated timing residuals from the Gaussian likelihood, $\tresid^{\textrm{sim}}\sim p(\tresid | \bm\bvec^{\textrm{sim}})$.

Each dataset $\tresid^{\textrm{sim}}$ is analyzed with the standard model that assumes a GW signal with a power-law spectrum. The only quantity that the predictive tests rely on is $p(\bm\Lambda |\tresid^{\textrm{sim}})$, i.e., the posterior for the hyperparameters, which we estimate through stochastic sampling with \textsc{Enterprise}~\cite{enterprise}. For computational efficiency, we ignore Hellings--Downs correlations during sampling as the posterior for the hyperparameters is dominated by the autocorrelation terms~{\cite{NANOGrav:2020spf,Romano:2020sxq,Hourihane:2022ner,Lamb:2023jls}. 

\section{Predictive checks on hyperparameters}
\label{sec:pch}

The most straightforward posterior predictive test performs comparisons directly at the level of the hyperparameters $\bm\Lambda$. In practise, this entails analyzing subsets of the data, for example by splitting the data of one PTA into two parts, or by analyzing data from one PTA only. The inferred GW amplitude and spectral slope are then used to predict the properties of the remaining data. 
However, given that current datasets are merely on the brink of making detections, splitting the data on one PTA will likely yield two uninformative datasets. 

Such predictive tests are related to consistency tests that directly contrast results across different PTAs, for example the posterior comparisons between EPTA, PPTA, and NANOGrav~\cite{Antoniadis:2022pcn}. That comparison used of the Mahalanobis distance~\cite{mahalanobis_distance} for the $\sigma$ deviations between two  $>1$-dimensional distributions, and found at most a $2.6\sigma$ deviation between different PTAs. We do not consider such tests in this study any further, instead leaving them to future work.

\section{Predictive checks on model parameters}
\label{sec:results-parameters}

The second posterior predictive test is based on the model parameters, and specifically the Gaussian process coefficients $\avec$. The comparison of the predicted and the inferred coefficients allows us to evaluate the power-law assumption of Eqs.~\eqref{eq:gw_power_law_definition} and~\eqref{eq:red_noise_definition}, as well as the assumption that the spatial correlations between pulsars follows the Hellings--Downs curve. 

The \textit{inferred} Gaussian-process coefficients are simply the inferred coefficients of the data. Stated differently, they are the Gaussian-process coefficients conditioned on the observed residuals, under the hypermodel prior.
Given the full posterior for model and hypermodel parameters $p(\bm\Lambda,\bvec| \tresid)$, Eq.~\eqref{eq:hyper_parameter_posterior} marginalizes over the parameters $\bvec$ to obtain the posterior for the hyperparameters.
Here we instead marginalize over the hyperparameters (and the timing-model parameters) to obtain the posterior for the Gaussian-process coefficients of the stochastic processes,
\begin{align}
    p_{\rm inf}(\avec| \tresid) = \int d\bm\Lambda \, d\bm\epsilon\; p(\avec,\bm \epsilon | \bm\Lambda, \tresid)\,p(\bm\Lambda| \tresid)\,.
    \label{eq:bvec-observed}
\end{align}
The first term in the integral is the posterior on $\bvec=[\epsvec, \avec]$  conditioned on both the timing residuals (i.e., the data $\tresid$) and the hyperparameters $\bm\Lambda$. In other words, $p_{\rm inf}(\avec | \tresid)$ is the posterior of the Gaussian-process coefficients under the hyperprior assumption that the observed data are subject to a common stochastic process and (optionally) Hellings--Downs-induced correlations from the inferred GW background.\footnote{In certain cases, stochastic sampling might yield the full posterior $p(\bvec,{\bm \Lambda}|\tresid)$, in which case $p(\avec | \tresid)$ can be obtained by marginalizing over $\bm \Lambda$ and $\bm \epsilon$. This is typically not the case for PTA analyses that sample from the marginalized posterior of Eq.~\eqref{eq:hyper_parameter_posterior}, we therefore have to reconstruct $p(\avec | \tresid)$ using Eq.~\eqref{eq:bvec-observed}.} 

The \textit{predicted} coefficients instead are only conditioned on the hyperparameter posterior, and not on the data directly:
\begin{align}
    p_{\rm pre}(\avec| \tresid) &= \int  d\bm\Lambda \, d\bm \epsilon\; p(\avec,\bm \epsilon | \bm\Lambda) p(\bm\Lambda | \tresid)\nonumber\\
    &= \int  d\bm\Lambda \; p(\avec | \bm\Lambda) p(\bm\Lambda | \tresid)\,.\label{eq:bvec-predicted}
\end{align}
Compared to Eq.~\eqref{eq:bvec-observed}, the first term in the integral is \textit{not} conditioned on $\tresid$.

The various terms in the integrands of Eqs.~\eqref{eq:bvec-observed} and~\eqref{eq:bvec-predicted} can be computed as follows. The hyperparameter posterior $p(\bm\Lambda |\tresid)$ is obtained by stochastic sampling via the analysis described in Sec.~\ref{sec:pta_likelihood}. The Gaussian-process coefficients conditioned on the hyperparameters are, by definition, given by a simplification of Eq.~\eqref{eq:model_parameters_hyper_prior}
\begin{equation}
    p(\avec | \bm \Lambda) = \frac{\exp\left(-\frac{1}{2}\avec^T \phimat^{-1}(\bm \Lambda)\avec\right)}{\sqrt{\det(2\pi\phimat(\bm \Lambda))}}\label{eq:gp_coefficients_hyper_prior}\,.
\end{equation}
The Gaussian-process coefficients and timing parameters conditioned on the hyperparameters and the data are
\begin{align}
     p(\avec, \bm \epsilon| \bm\Lambda, \tresid)= \frac{p(\tresid | \avec, \bm \epsilon, \bm\Lambda) p(\avec, \bm \epsilon | \bm\Lambda)}{p(\tresid | \bm\Lambda)}=\mathcal N(\widehat\bvec, \bm\Sigma)\label{eq:gp_coefficients_conditioned}\,,
\end{align}
where in the first equality we have used Bayes' theorem and $\mathcal N(\widehat \bvec,\bm\Sigma)$ indicates a normal distribution with mean and covariance given by Eqs.~\eqref{eq:bvec_max_likelihood} and~\eqref{eq:bvec_covariance_matrix}.

To construct the predicted coefficients we sample Eq.~\eqref{eq:bvec-predicted} by first drawing $\bm\Lambda^s \sim p(\bm\Lambda | \tresid)$, then using the sample $\bm\Lambda^s$ to construct $\phimat^s(\bm \Lambda)$ and draw from Eq.~\eqref{eq:gp_coefficients_hyper_prior}.
The amplitude of these coefficients should, on average, be consistent with the assumed power-law model\footnote{This is true if $\gamma$ for the power-law model is fixed. If the spectral index is sampled over then the power reconstructed from an individual draw for $\avec$ will, on average, be consistent with a power-law associated with the $\gamma$ for that specific draw.}.
To construct the inferred coefficients we sample Eq.~\eqref{eq:bvec-observed} by first drawing $\bm\Lambda^s \sim p(\bm\Lambda | \tresid)$, then using the sample $\bm\Lambda^s$ to construct $\phimat^s(\bm \Lambda)$, $\widehat\bvec^s$, and $\bm\Sigma^s$ and draw from Eq.~\eqref{eq:gp_coefficients_conditioned}.
The amplitude of the inferred coefficients has a power-law hyperprior, but is also conditioned on the data and can this deviate from a pure power-law.

Besides the assumption of a power-law common process, we can further use the inferred and predicted distributions to test the nature of the spatial correlations.
Both Eqs.~\eqref{eq:gp_coefficients_hyper_prior} and~\eqref{eq:gp_coefficients_conditioned} depend on $\phimat(\bm \Lambda)$, whose non-diagonal terms encode the inter-pulsar correlations. 
We can therefore evaluate the inferred and predicted distributions by assuming a correlation pattern, such as Hellings--Downs or monopolar correlations.
On average, the predicted coefficients will have the assumed correlation pattern. The inferred coefficients will have a correlation pattern informed by the data, but subject to the hyperprior of a power-law common process with the assumed correlation pattern.
A discrepancy between these predicted and inferred distributions would signal that the assumed pattern is not consistent with the data. In this work, we focus on visual discrepancies that can be seen from the figures, however, one could also consider constructing associated $p$-values~\cite{ppc1}.

\subsection{Intrinsic noise model}

\begin{figure}
    \centering
    \includegraphics[width=0.49\textwidth]{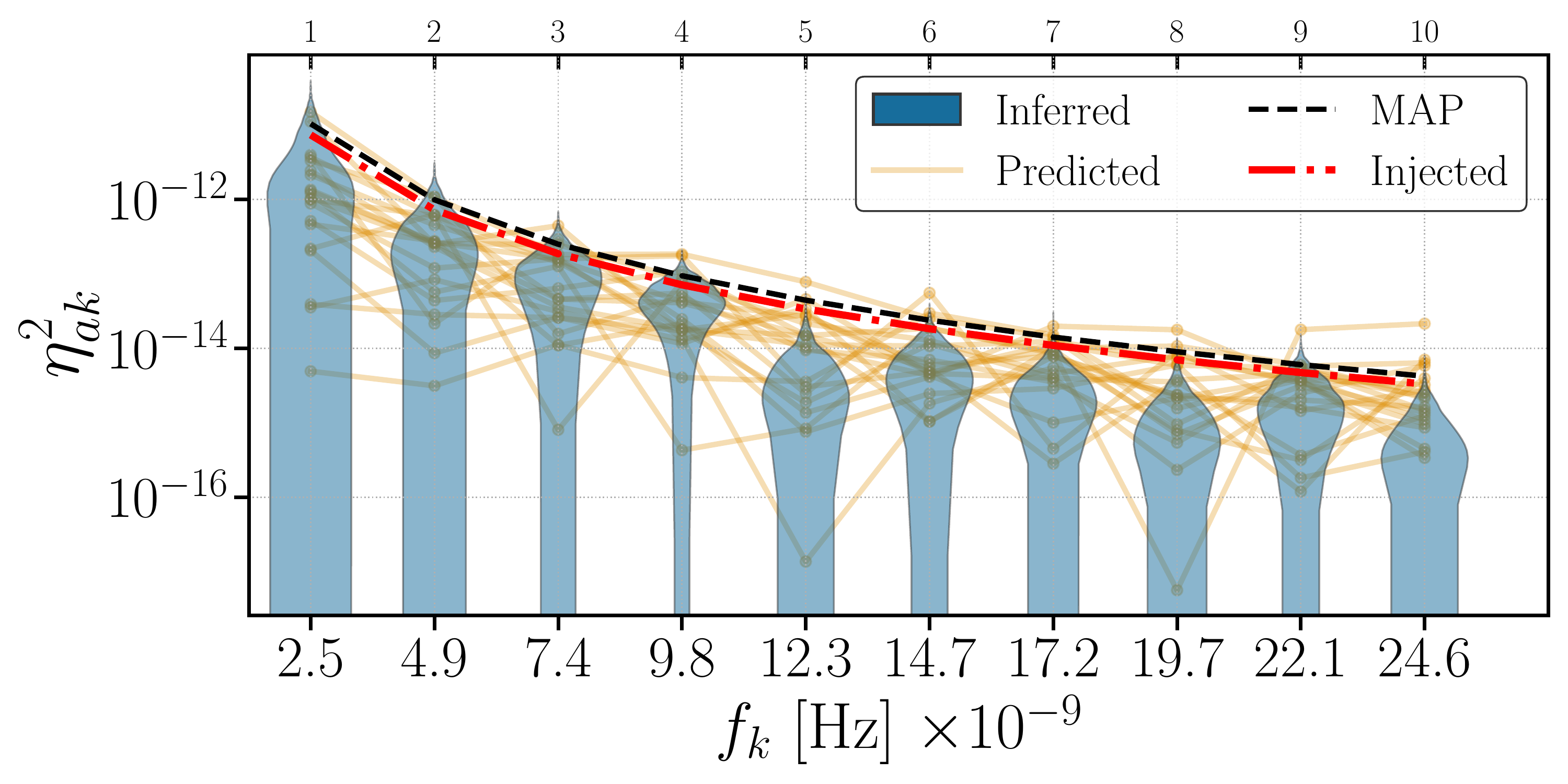}
    \caption{Intrinsic pulsar noise power, $\eta_{ak}^2$ from Eq.~\eqref{eq:red_noise_definition}, as a function of frequency $f_k$ (bottom x-axis, $k$ index on the top x-axis) for B1937+21 for the \textsc{HellingsDowns-PowerLaw} simulated dataset. Sample predicted power spectra are shown in orange. The blue violins show the posterior for the inferred power at each frequency, which is a combination of the data and the power-law prior. For reference, we plot the injected and maximum \textit{a posteriori} power-law spectra in red dot-dashed and black dashed lines respectively.
    }
    \label{fig:red_noise_posteriors}
\end{figure}

We begin by applying the above methodology to pulsar intrinsic noise, which is modeled with Eq.~\eqref{eq:red_noise_definition}. The relevant model parameters are the sine and cosine amplitudes associated with each frequency, $a^{(s)}_{i,a}$ and $a^{(c)}_{i,a}$ respectively for pulsar $a$ and frequency bin $i$.
Specifically, we use Eqs.~\eqref{eq:gp_coefficients_hyper_prior} and~\eqref{eq:gp_coefficients_conditioned} to draw from the inferred and predicted distribution of the intrinsic noise in pulsar $a$ and frequency bin $i$ and then obtain the total power as the square-sum of the sine and cosine components,
\begin{equation}    \eta_{ai}^2=\frac{1}{2}\left\{\left[a^{(s)}_{i,a}\right]^2+\left[a^{(c)}_{i,a}\right]^2\right\}\,.
\end{equation}
Each of $a^{(s)}_{i,a}$ and $a^{(c)}_{i,a}$ is normally distributed according to the intrinsic-pulsar-noise power spectrum, Eq.~\eqref{eq:gp_coefficients_hyper_prior}, so the total power at each frequency follows a $\chi^2$ distribution with $2$ degrees of freedom for a given $\bm\Lambda^s$.

Results for a representative pulsar are shown in Fig.~\ref{fig:red_noise_posteriors} using the \textsc{HellingsDowns-PowerLaw} simulated dataset. 
We show inferred (blue) and predicted (orange) spectra as a function of frequency. For reference, we also show the injected and maximum \emph{a posteriori} spectrum. The inferred power is only significantly constrained away from zero at the fourth frequency bin, while the predicted power are wider. In most bins, the inferred and predicted distributions have comparable width (given the logarithmic scale on the $y$ axis), suggesting that the data are not strongly informative.
The inferred and predicted distributions overlap for all frequencies, as expected since the simulated dataset includes intrinsic noise that obeys the power-law assumption.

\subsection{ GW-background model}

We now turn our attention to arguably the most important part of the analysis: the GW background. Detection of the GW background hinges on establishing that the data follow the Hellings--Downs correlation pattern, while the astrophysical interpretation of the signal relies on its spectral shape, specifically the amplitude and slope of the assumed power-law~\cite{Sampson:2015ada,Taylor:2016ftv,Middleton:2020asl,Becsy:2022pnr}. Below we apply posterior predictive checks to assess both elements.

\subsubsection{GW power spectrum of individual pulsars}

\begin{figure*}
    \centering
    \includegraphics[width=0.49\textwidth] {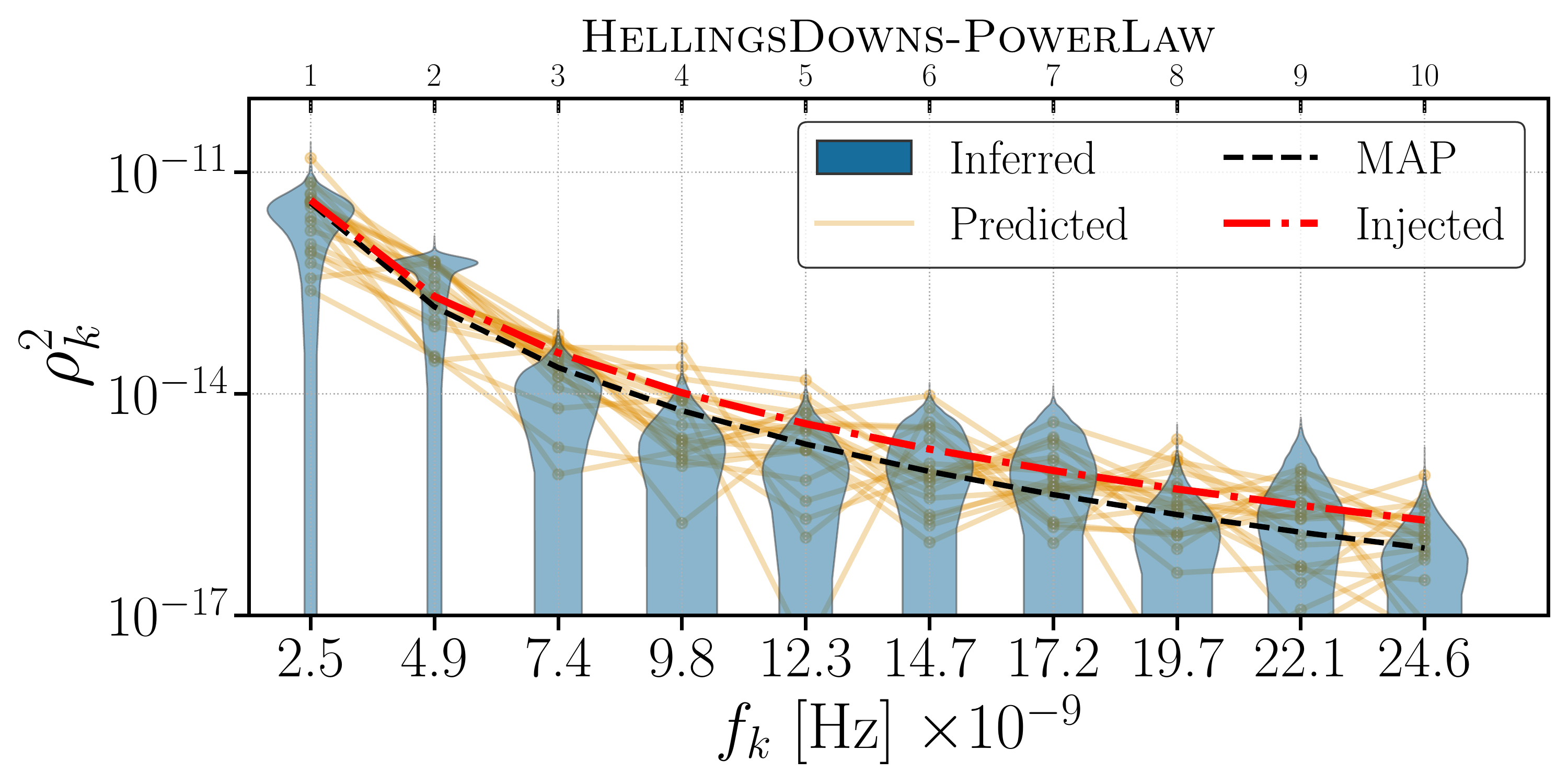}    \includegraphics[width=0.24\textwidth]{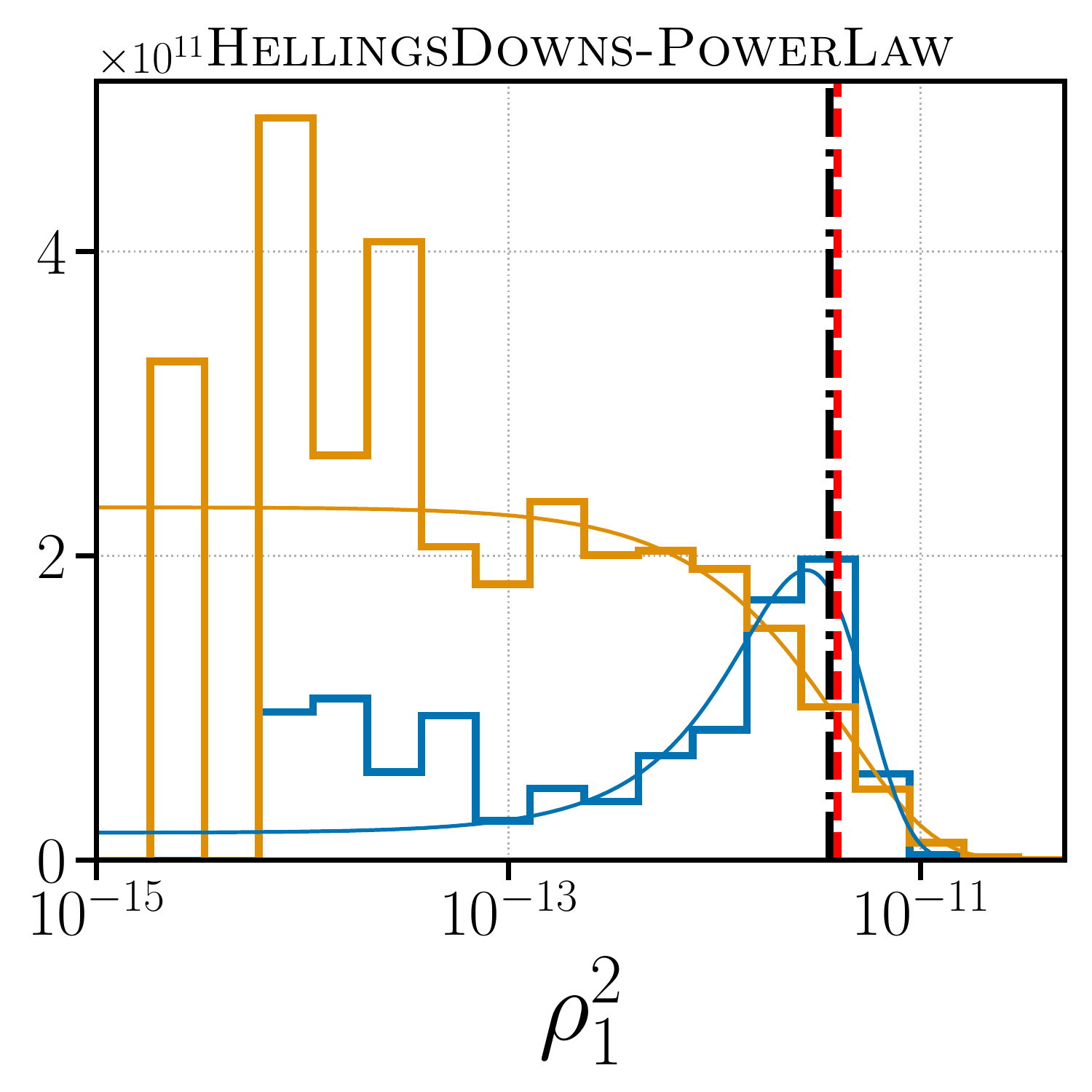}
    \includegraphics[width=0.24\textwidth]{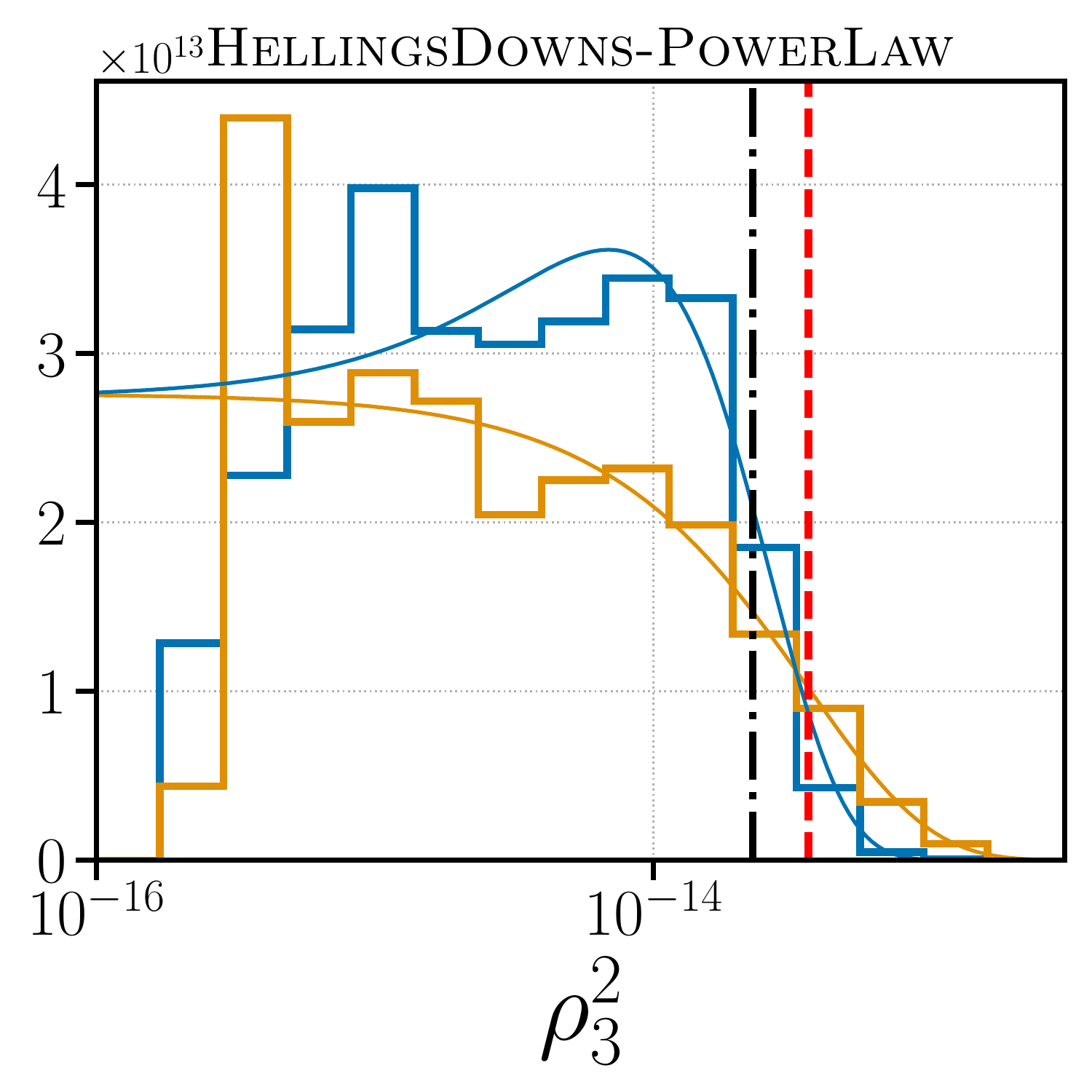}\\
    \includegraphics[width=0.49\textwidth]{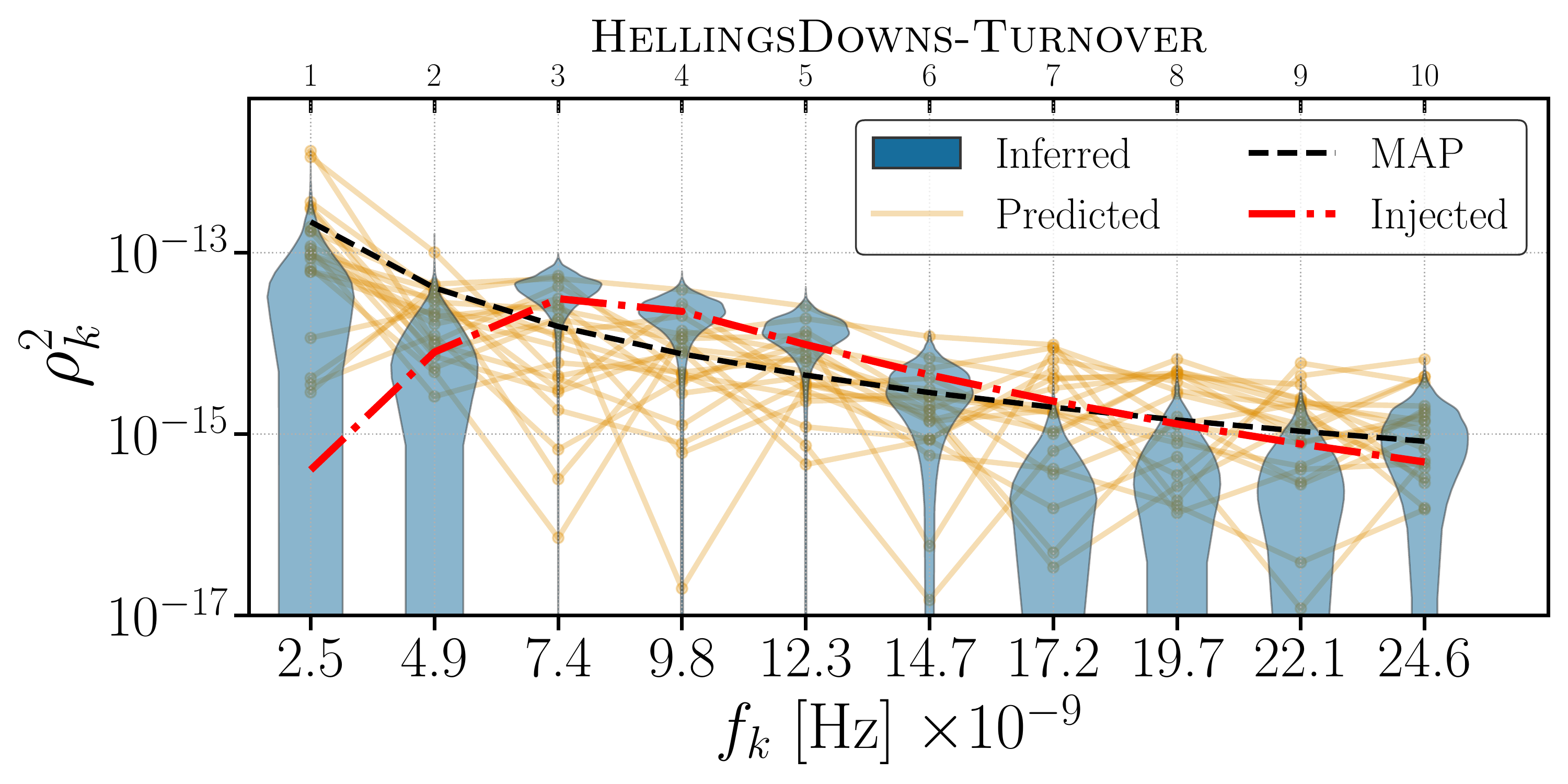}
    \includegraphics[width=0.24\textwidth]{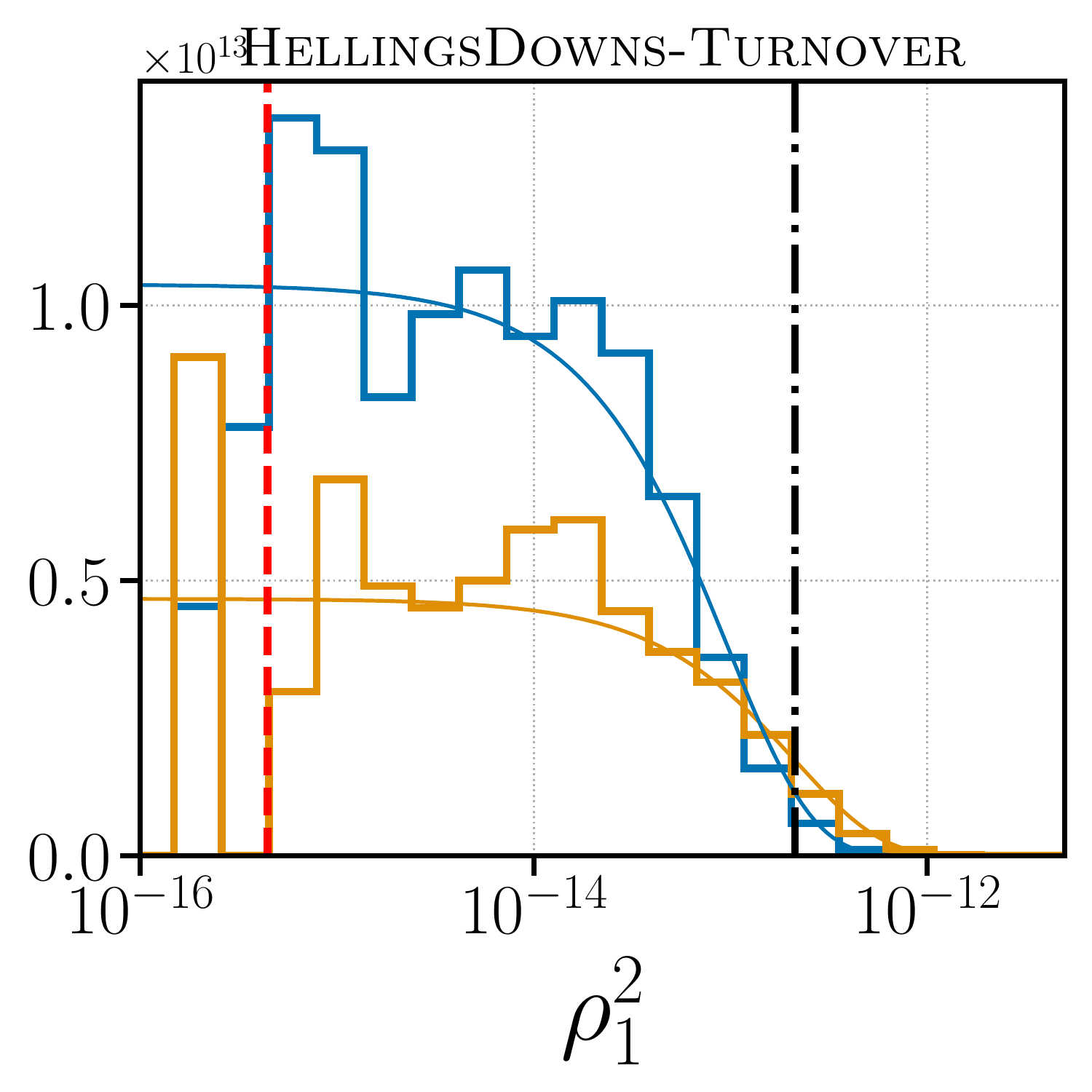}
    \includegraphics[width=0.24\textwidth]{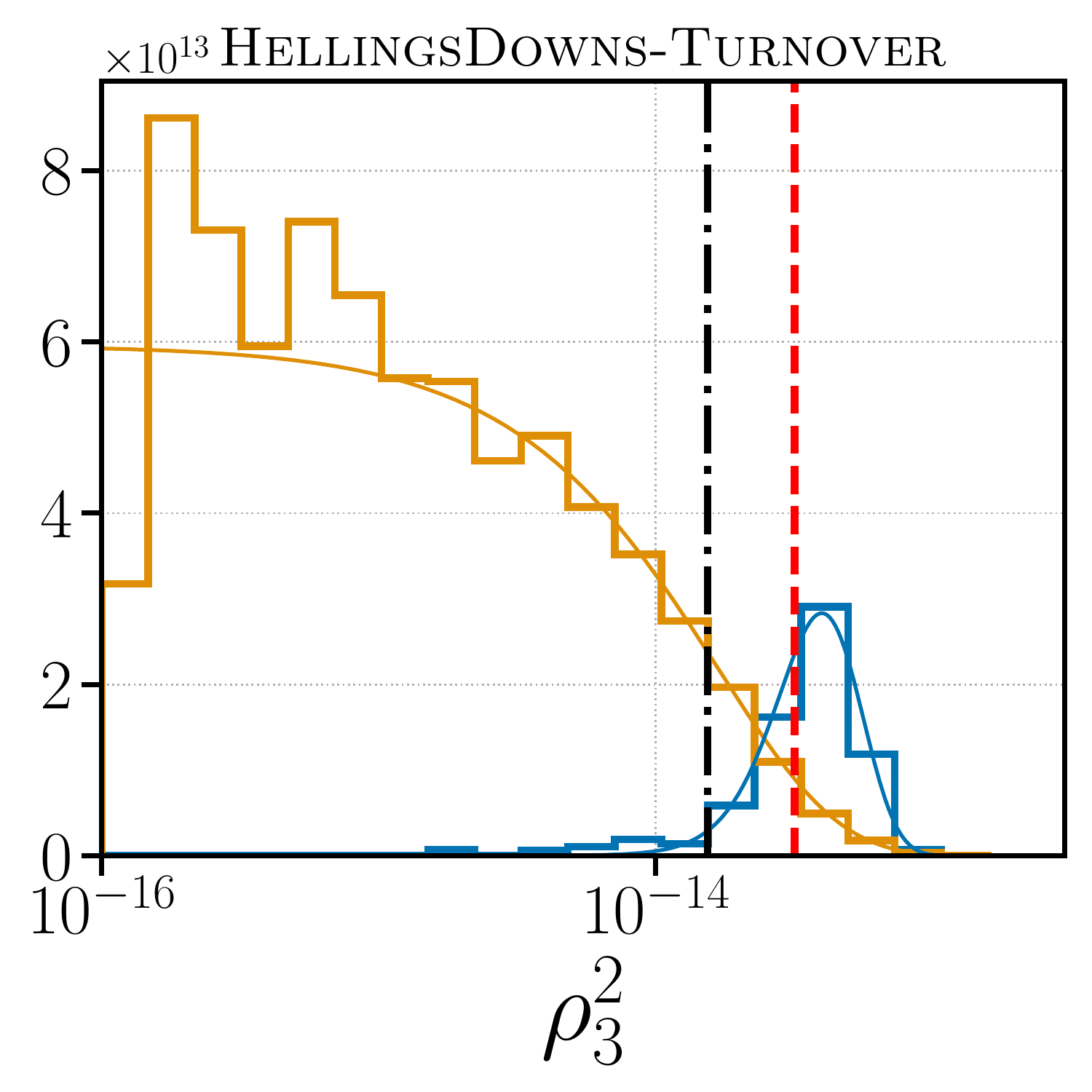}
    \caption{GW power spectrum, $\rho^2_{k}$ from Eq.~\eqref{eq:gw_power_law_definition}, as a function of the frequency $f_k$ (left) and power distributions for select frequency bins (right) for an ``informative'' pulsar, J1909--3744,
    for the \textsc{HellingsDowns-PowerLaw} (top) and the \textsc{HellingsDowns-Turnover} (bottom) dataset. 
    In the left panels, sample predicted power spectra are shown in orange and blue violins show the posterior for the inferred power at each frequency. For reference, we plot the injected and maximum \textit{a posteriori} spectra in red dot-dashed and black dashed lines respectively. 
    In the right panels, we show histograms of the inferred and predicted power for the $1^\mathrm{st}$ and $3^\mathrm{rd}$ bins, along with a fit to a $\chi^2$ distribution with two degrees of freedom. 
    In the top panels, we find agreement between the predicted and inferred spectra for the data-informed frequency bins, i.e., the ones constrained away from zero. In the bottom panel, data-informed  bins contain systematically higher power than the prediction, as expected from the injected spectra. }
    \label{fig:pulsar_gwb_realizations}
\end{figure*}

While the GW background has a single power spectrum across all pulsars as in Eq.~\eqref{eq:gw_power_law_definition}, the exact realization in each pulsar is unique\footnote{This is in part, but not solely, due to the ``pulsar term" that Hellings--Downs correlations do not capture.}, and this results in different Gaussian process coefficients.
We therefore begin by considering the inferred and predicted GW power in individual pulsars. 
Figure~\ref{fig:pulsar_gwb_realizations} shows power spectra (left) and power distributions for frequency bins of interest (right) for an ``informative" pulsar with detectable GW power in some bins.
The top panels show results for the \textsc{HellingsDowns-PowerLaw} dataset, while the bottom panels correspond to \textsc{HellingsDowns-Turnover}. Both datasets are analyzed with the same GW model, hence the maximum \emph{a posteriori} draw and the predicted spectra are power-laws.   

The posterior predictive test proceeds as follows. First, we analyze the data assuming a power-law model and (inevitably) infer power-law parameters that fit the data as well as possible. The predicted spectra are draws from this inferred power-law. The maximum \emph{a posteriori} draw is essentially the power-law model's best attempt to match the true spectrum.  Second, the inferred spectra are the power in the data inferred under a GW spectrum prior that is the inferred power-law posterior. The final inferred spectra are thus a combination of the data and the prior. For informative pulsars, in a few of the frequency bins the data dominate over the power-law prior. For uninformative pulsars on, on the other hand, the inferred spectra would be consistent with the power-law imposed by the prior in all bins. 

Indeed, in Fig.~\ref{fig:pulsar_gwb_realizations} the $1^\mathrm{st}$--$2^\mathrm{nd}$ (top) and $3^\mathrm{rd}$--$6^\mathrm{th}$ (bottom) frequency bins have inferred spectra that are constrained away from zero. The inferred spectra in these bins are narrower than the predicted ones, suggesting informative data. In the top panel the inferred and predicted spectra fully overlap since the model matches the simulated spectrum. In the bottom panel, however, the
inferred spectra are systematically higher than the predicted ones. Moreover, the $1^\mathrm{st}$--$2^\mathrm{nd}$ bins are consistent with zero, which is in tension with expectations from a power-law. 
This behavior is due to the fact that the injection follows a power-law with a turnover, which the GW power-law model cannot fully match, as manifest in the maximum \emph{a posteriori} draw. The inferred spectra are therefore dominated by the data and reveal a tension with the predicted spectra. 

Though not explicitly plotted, we have verified that for uninformative pulsars, i.e., pulsar with high intrinsic noise with no detectable GW power, the inferred and predicted distributions are nearly identical. This suggests that the total inference is dominated by the prior. 

\subsubsection{Total GW power spectrum}

In order to obtain an estimate of the total GW power spectrum, we use the optimal statistic~\cite{Anholm:2008wy,2015PhRvD..91d4048C,Vigeland:2018ipb}, which is based on the 
timing residuals from all pulsars. The optimal statistic gives a noise-weighted average of the cross-correlation between pulsar pairs, and therefore allows us to synthesize the inferred or predicted coefficients from different pulsars into a single estimate of the GW background amplitude. Since we are testing the GW model, we reconstruct the optimal statistic using only the GW contribution to the timing residuals and ignore the timing model and intrinsic pulsar noise parts.

We obtain draws for the Gaussian process coefficients $\avec^s$ of the GW background through Eqs.~\eqref{eq:bvec-observed} or~\eqref{eq:bvec-predicted} as applicable, and construct timing residuals $\tresid^s = \fmat\avec^s$.
We then use the optimal statistic to compute inter-pulsar cross-correlations $\xi^s_{ab,k}$ and GW background amplitude $A^s_{\mathrm{gw}}$ for each frequency bin $k$.\footnote{This ``per-frequency" optimal statistic as compared to the most common summed-over-frequencies version is studied in~\cite{OScorr}.} For a pair of pulsars $a$ and $b$, the former is
\begin{align}
    \xi_{ab,k} &= \frac{\tresid_a^T\dmat_a^{-1}\tilde \Phimat_{ab,k}^{\mathrm{gw}}\dmat_b^{-1}\tresid_b}{\mathrm{tr}\left(\dmat_a^{-1}\tilde\Phimat_{ab,k}^{\mathrm{gw}}\dmat_b^{-1}\tilde\Phimat_{ba,i}^{\mathrm{gw}}\right)}\label{eq:os_single_corr_single_bin}\,,\\
    \sigma_{ab,k}^2 &= \left[\mathrm{tr}\left(\dmat_a^{-1}\tilde\Phimat_{ab,k}^{\mathrm{gw}}\dmat_b^{-1}\tilde\Phimat_{ba,i}^{\mathrm{gw}}\right)\right]^{-1}\label{eq:os_sigma_single_corr_single_bin}\,,
\end{align}
where no summation is implied. In the above equations we have defined $\tilde\Phimat_{ab,k}^{\mathrm{gw}}=\fmat_{a,k}\tilde\phimat_{ab,k}^{\textrm{gw}}\fmat^T_{b,k}$ where
\begin{align}
\tilde{\phimat}_{ab,k}^{\textrm{gw}}= \Gamma_{ab}\frac{1}{12\pi^2} \frac{f_\textrm{y}^{-3}}{T}\,,\label{eq:phi_mat_binned_estimator}
\end{align}
is a GW-only normalized version of Eq.~\eqref{eq:phi_matrix_definition}. The subscripts in $\fmat_{a,k}$ denote that it is evaluated at the times for which pulsar $a$ has data and for \textit{only} frequency $k$. The matrix $\dmat_a = [\cmat(\bm\Lambda)]_{(ai,aj)}$ is the autocorrelation block for pulsar $a$ of the marginalized covariance matrix used in Eq.~\eqref{eq:hyper_parameter_posterior}, and depends on the hyperparameters $\bm\Lambda$. It represents the total noise autocorrelation for pulsar $a$ from both uncorrelated and correlated processes. The normalization in Eq.~\eqref{eq:phi_mat_binned_estimator} is chosen such that $\xi_{ab, k}$ is an estimator for the GW background in each frequency bin.

Given $\xi_{ab,k}$ we construct a bin-by-bin estimator for the GW background obtained through a weighted average across all pulsar pairs, 
\begin{align}
    \xi_k &= \frac{\sum_{ab}\xi_{ab,k}\sigma_{ab,k}^{-2}}{\sum_{ab} \sigma_{ab,k}^{-2}}\,,\label{eq:final_bin_by_bin_estimator}\\
   \sigma_{k}^2 &=  \left[\sum_{ab} \sigma_{ab,k}^{-2}\right]^{-1}\,\label{eq:final_bin_by_bin_estimator_sigma}.
\end{align}
These equations assume independent frequency bins and pair correlations, which is not strictly true~\cite{OScorr}. In the weak-GW limit, 
the frequency bins and paired correlations are approximately uncorrelated,  but for strong signals such as those that we inject here the covariances between pair correlations become significant \cite{Allen:2022dzg,Allen:2022ksj,Bernardo:2022xzl,Bernardo:2023bqx,NANOGrav:2023icp,OScorr}.
We nevertheless ignore them in this work for the sake of computational efficiency. Including them would broaden the green and blue violins for both the spectral and correlation reconstructions in Figures~\ref{fig:total_gwb_spectra} and ~\ref{fig:angular_correlations}~\cite{OScorr}.

\begin{figure*}
    \centering    \includegraphics[width=0.49\textwidth]{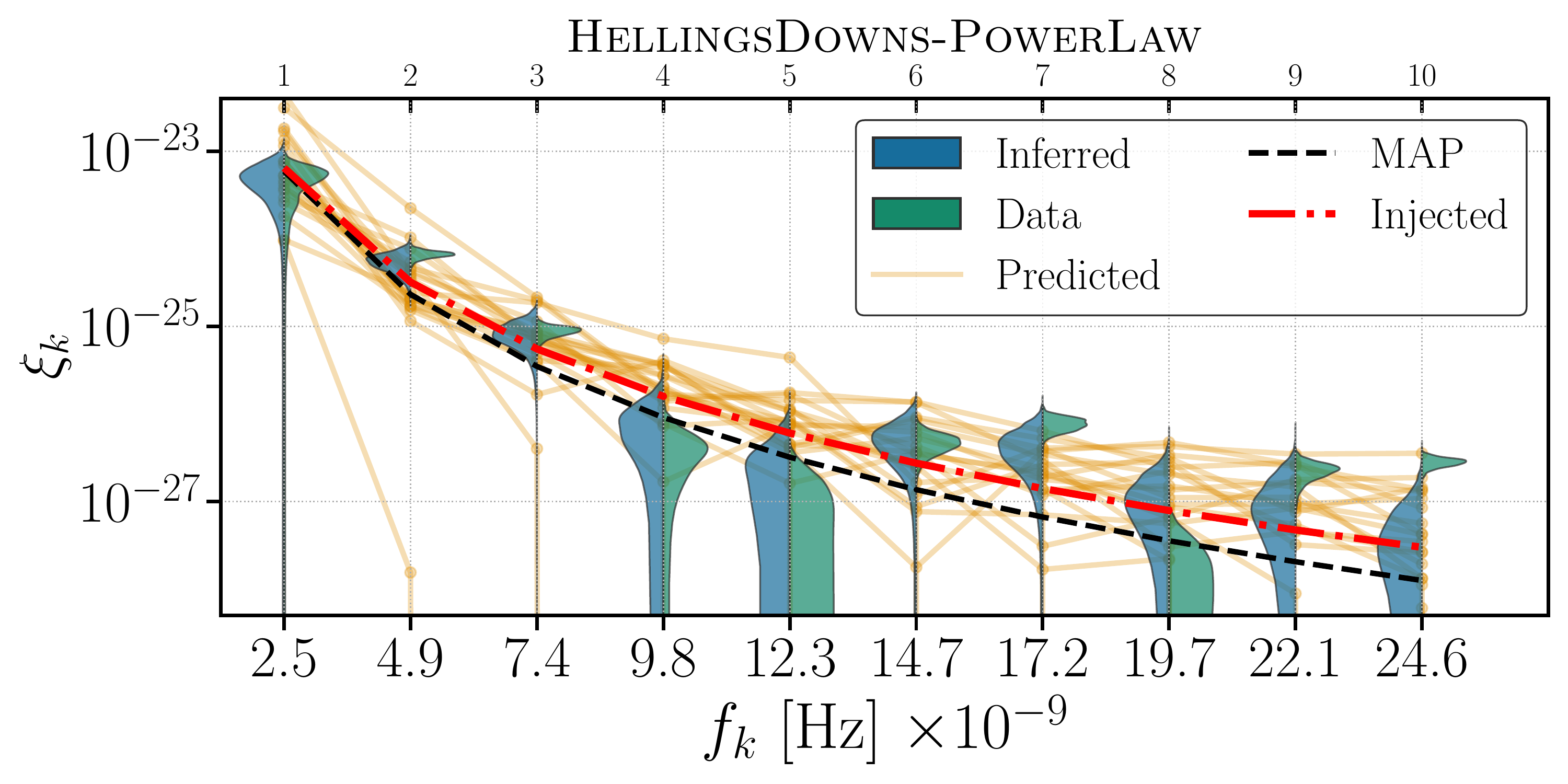}
    \includegraphics[width=0.24\textwidth]{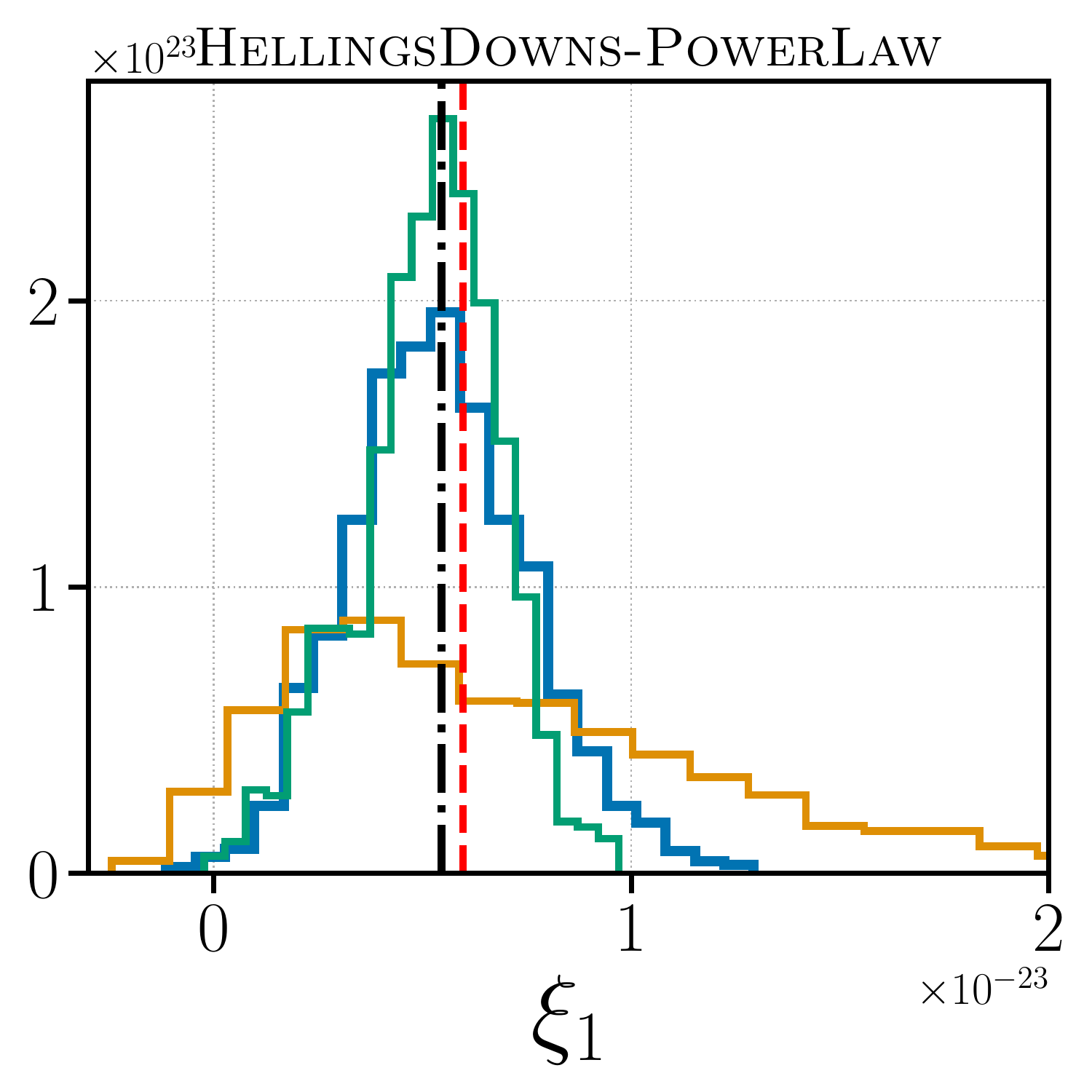}
    \includegraphics[width=0.24\textwidth]{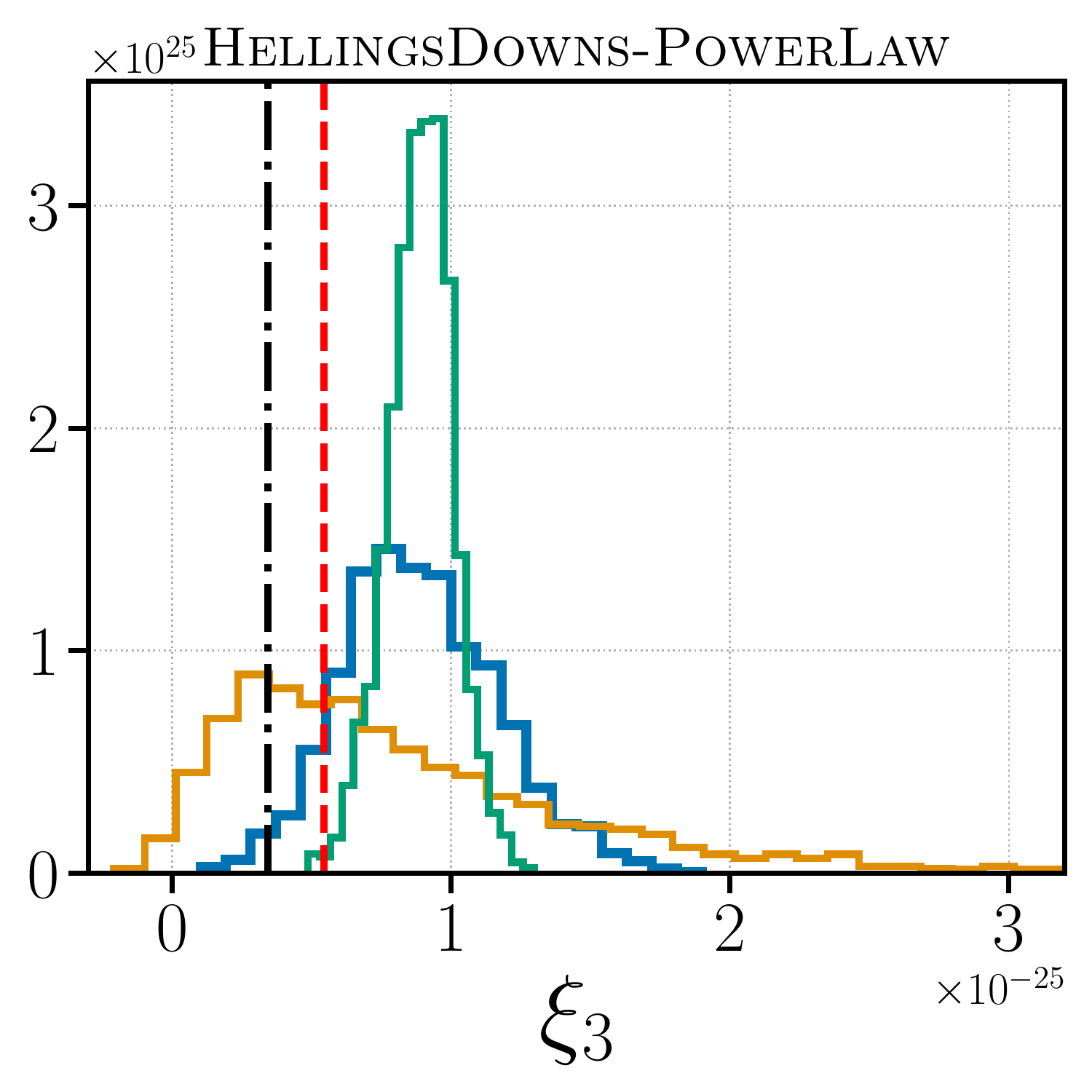}\\
    \includegraphics[width=0.49\textwidth]{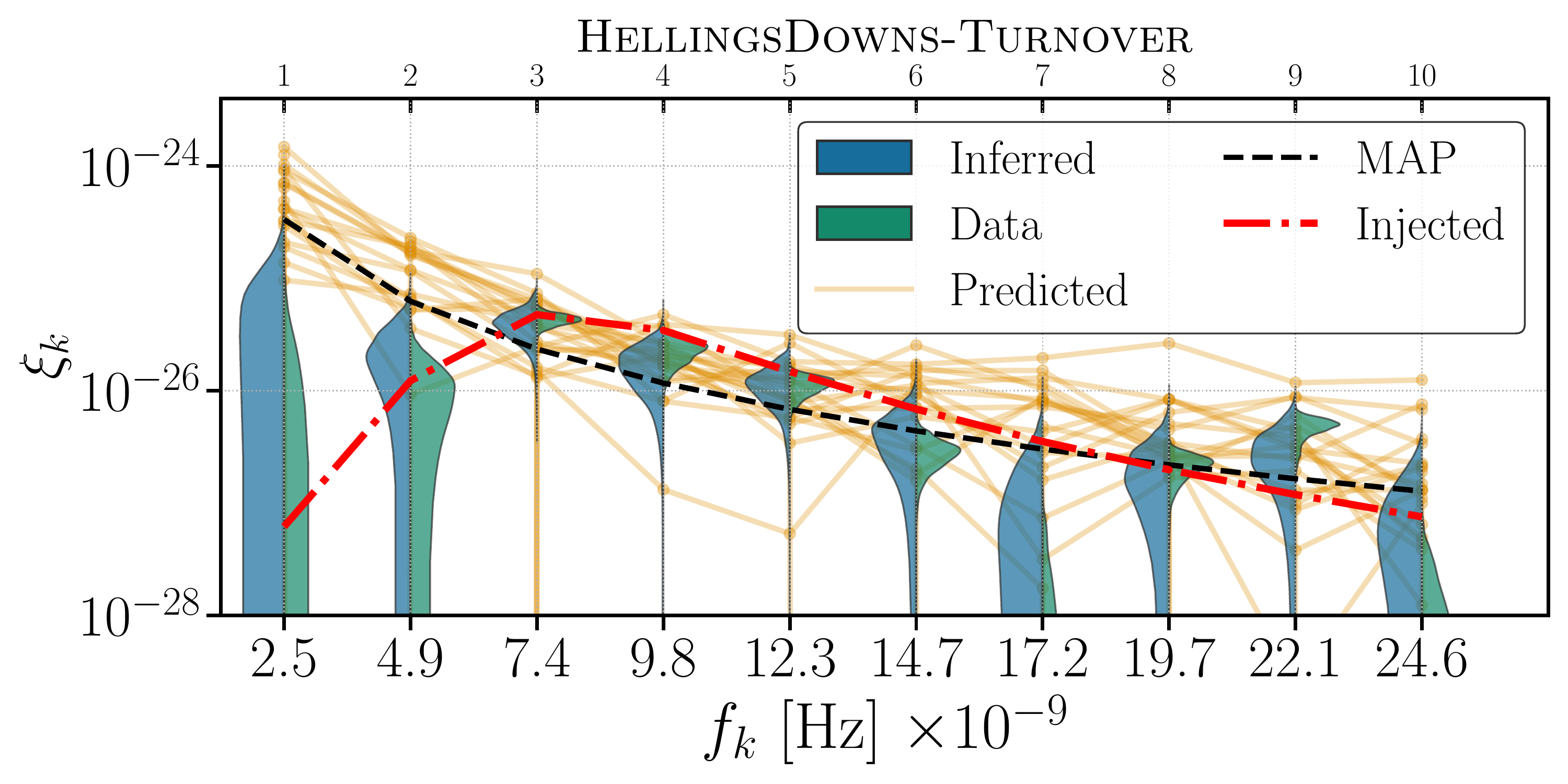}
        \includegraphics[width=0.24\textwidth]{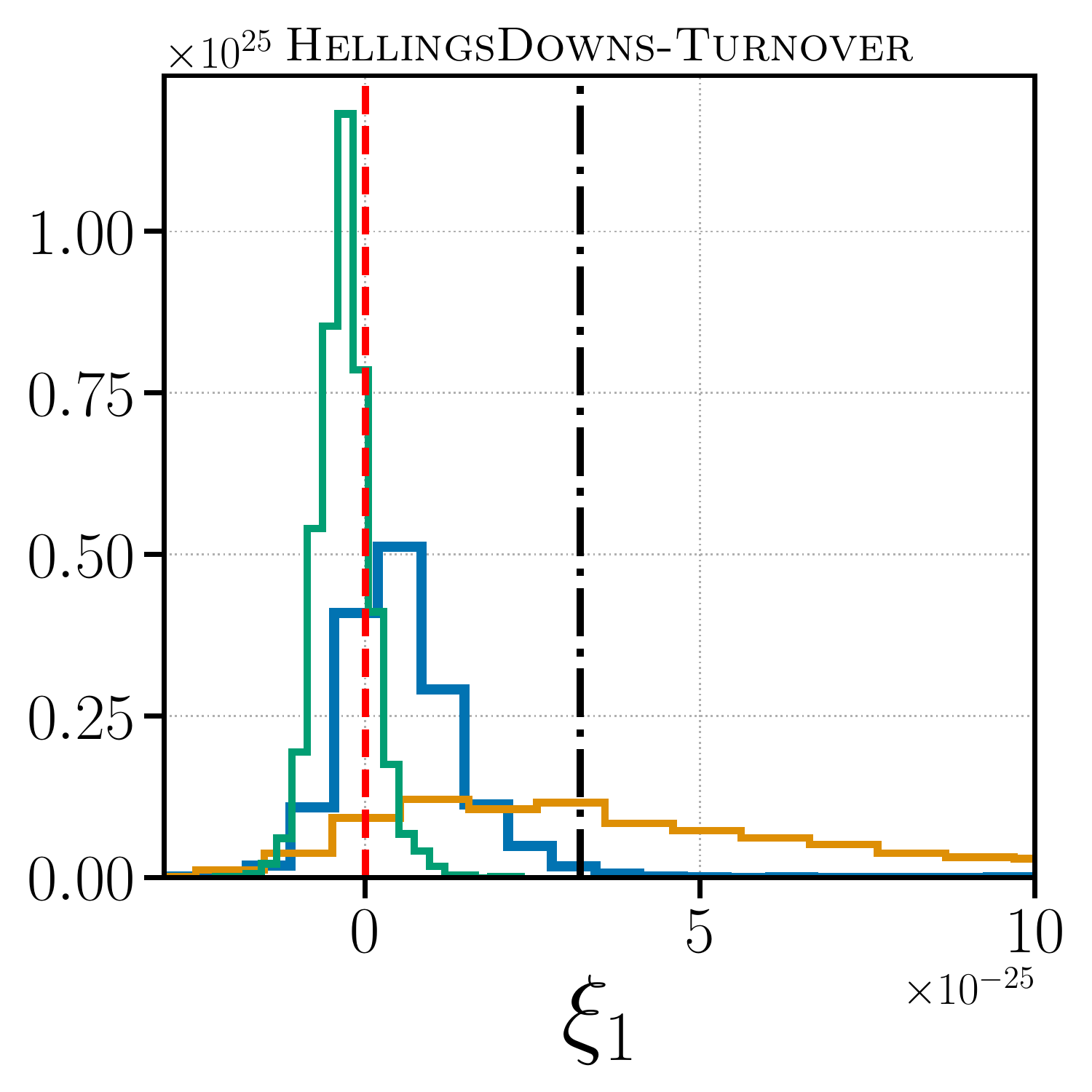}
    \includegraphics[width=0.24\textwidth]{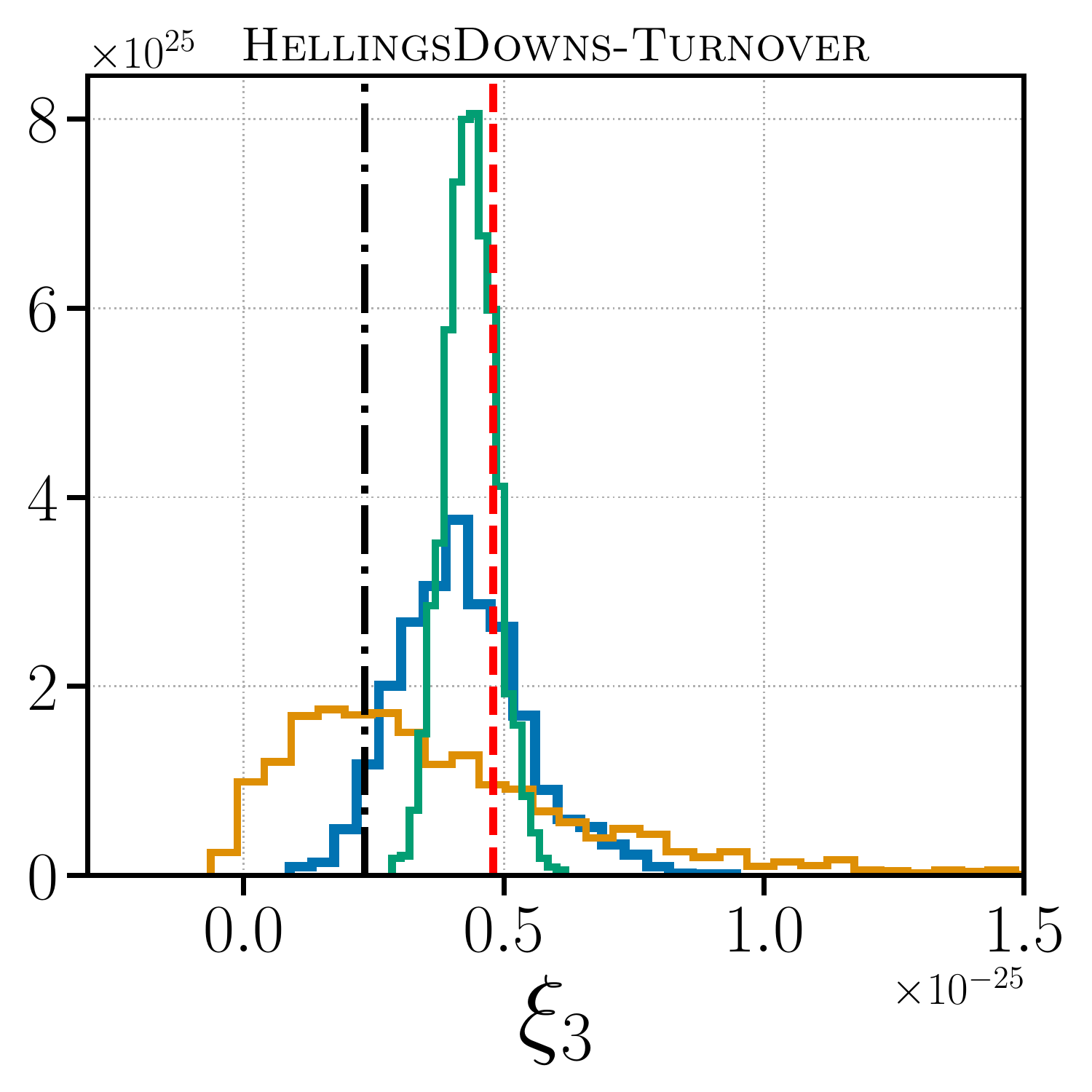}
    \caption{Total GW power spectrum, $\xi_k$ from Eq.~\eqref{eq:final_bin_by_bin_estimator}, as a function of frequency $f_k$ (left) and total power distributions for select bins (right) for the \textsc{HellingsDowns-PowerLaw} (top) and \textsc{HellingsDowns-Turnover} (bottom) datasets. 
    In the left panels we show the inferred (blue left violins) and predicted (orange lines) distributions using the single-frequency optimal statistic. The injected and maximum \textit{a posteriori} power-law spectra are shown in red dot-dashed and black dashed lines respectively. Right green violins show the power as inferred directly from the data without conditioning on a power-law spectrum. 
     In the right panels, we show histograms of the inferred and predicted power for the $1^\mathrm{st}$ and $3^\mathrm{rd}$ bins, along with a fit to a $\chi^2$ distribution with two degrees of freedom.
    Inferred and predicted spectra are consistent in the top panel. However, the inferred power in the $1^{st}$ and $2^{nd}$ frequency bins in the bottom panel is lower than what predicted under the power-law model.
    }
    \label{fig:total_gwb_spectra}
\end{figure*}

Figure~\ref{fig:total_gwb_spectra} shows the total GW spectrum (left) and power distributions for select bins (right) for the \textsc{HellingsDowns-PowerLaw} (top) and \textsc{HellingsDowns-Turnover} (bottom) datasets. We present the same inferred, predicted, maximum \emph{a posteriori}, and injected spectra as in Fig.~\ref{fig:pulsar_gwb_realizations}. Additionally, we calculate Eqs.~\eqref{eq:final_bin_by_bin_estimator} and~\eqref{eq:final_bin_by_bin_estimator_sigma} directly using the original simulated data and obtain an estimate that is informed solely by the data without assumptions about the GW spectral shape. The various spectra represent the optimal statistic calculated on the predicted, inferred, and simulated data for the same set of posterior samples drawn from $p(\bm\Lambda | \tresid)$. For the inferred and predicted case, the hyperparameters are used to construct the GW coefficients $\avec^s$ and $\dmat_a$, while for the data, the hyperparameters are only needed in the construction of $\dmat_a$.
The predicted estimate corresponds to power-law spectra whose amplitude and slope have been inferred by the data. The inferred estimate is a combination of data and prior: it corresponds to the GW spectrum as observed by all pulsars and under the assumption of a power-law. Thus, the predicted estimate will always follow a power-law, while the inferred estimate will shift the spectra as close to a power-law as the data allow.

Starting with the top panel of Fig.~\ref{fig:total_gwb_spectra} and the \textsc{HellingsDowns-PowerLaw} dataset, we find that the predicted and inferred data on average overlap with some scatter. 
In places where the data contain higher power than the injected power-law, e.g., $6^\mathrm{th}$ and $7^\mathrm{th}$ frequency bins, the inferred estimate is wider and shifted down toward the power-law. In some cases, such as the 
$9^\mathrm{th}$ and $10^\mathrm{th}$ bins, what looks like a GW detection from the data turns out to be insignificant when estimated in the context of the power-law model. Despite these, for the most informative $1^\mathrm{st}$, $2^\mathrm{nd}$ and $3^\mathrm{rd}$ bins, the observed data fully agree with the power-law model as expected. 

Moving to the bottom panel of Fig.~\ref{fig:total_gwb_spectra} and the \textsc{HellingsDowns-Turnover} dataset, the spectra comparison is drastically different. The most significant bins are now the $3^{rd}$, $4^{th}$ and $5^{th}$ ones as expected from the injected spectrum shape. These bins agree with the predicted distribution, suggesting that they largely drive the inference of the power-law amplitude. However, the $1^{st}$ and $2^{nd}$ bin are consistent with no GW power and are systematically lower than the power-law model prediction. As expected, the inferred distribution is shifted upwards compared to the data-only distribution, attempting to match the power-law model. However, the data place strong upper limits on the GW power in those bins and the tension between the predicted and inferred distributions is apparent. 
\begin{figure}
    \centering
    \includegraphics[width=0.49\textwidth]{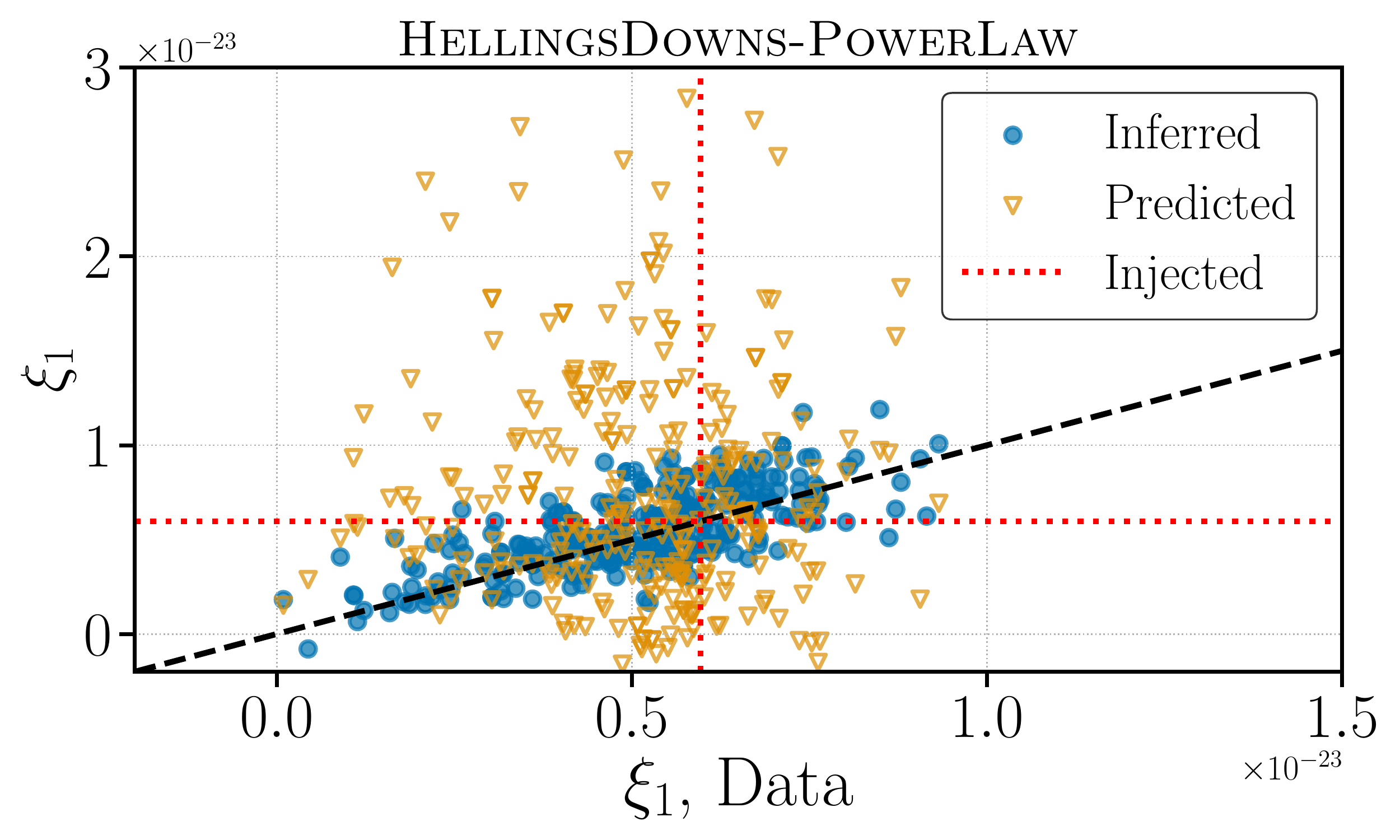}
    \includegraphics[width=0.49\textwidth]{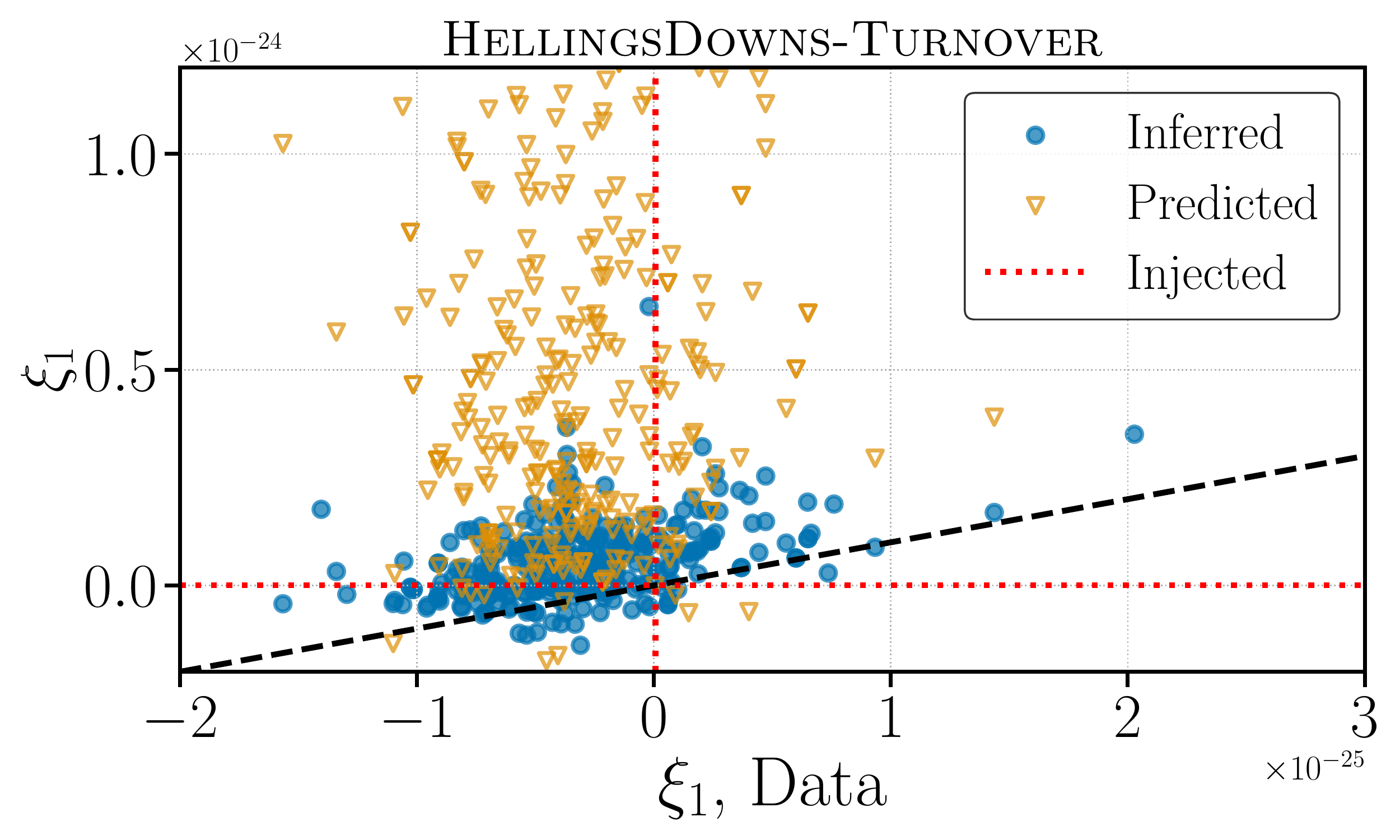}
    
    \caption{Scatter plot comparison of the power in the first frequency bin for the data only vs. the predicted (orange) and inferred (blue) power for the \textsc{HellingsDowns-PowerLaw} (top) and \textsc{HellingsDowns-Turnover} (bottom) datasets. Each point is a draw from the distributions shown in Fig.~\ref{fig:total_gwb_spectra}. In the top panel the bulk of the predicted and inferred draws overlap, while the inferred draws follows the $x-y$ lines as expected from highly informative data. In the bottom panel, the predicted draws overestimate the GW power.}
    \label{fig:frequency_bin_scatter_plots}
\end{figure}

Beyond the full distributions shown in Fig.~\ref{fig:total_gwb_spectra}, we compare the various spectra estimates on a draw-by-draw basis in Fig.~\ref{fig:frequency_bin_scatter_plots}. We show a scatter plot of $\xi_1$ for $300$ posterior draws from the \textsc{HellingsDowns-PowerLaw} (top) and \textsc{HellingsDowns-Turnover} (bottom) datasets. The $x$-axis shows the value calculated on the measured data, while the $y$-axis shows the predicted and inferred $\xi_1$. In the top panel, inferred draws are narrower than predicted draws and stay close to the $x-y$ line, an outcome of the fact that the data are very informative in this bin. In the bottom panel the inferred draws are more weakly correlated with the data draws, and shifted upward due to the power-law prior. Additionally, the bulk of the predicted draws overlap with the inferred ones in the top panel, which we expect because the model used for the predicted draws matches the injected model.  In the bottom panel the predicted draws have a larger tail toward higher values, as the power-law model overestimates the GW power in this frequency bin.

\subsubsection{Spatial correlations}

The predicted and inferred data can also be compared to assess consistency with the Hellings--Downs correlation pattern. We correlate data between pulsars using the full frequency band version of Eq.~\eqref{eq:os_single_corr_single_bin}, i.e., we use the full $\phimat_{ab}^{\textrm{gw}}$ instead of $\phimat_{ab,k}^{\textrm{gw}}$, so we drop the subscript $k$ and write $\xi_{ab}$. Additionally, since the Hellings--Downs model is already built in to the optimal statistic, we divide Eq.~\eqref{eq:os_single_corr_single_bin} by $\Gamma_{ab}$ and Eq.~\eqref{eq:os_sigma_single_corr_single_bin} by $\Gamma_{ab}^2$. We denote these ``normalized" correlations with $\tilde\xi_{ab} \equiv \xi_{ab}/\Gamma_{ab}$. Finally, we collect the $\tilde\xi_{ab}$'s into $8$ bins (each containing approximately the same number of pulsar pairs) based on the pair angular separation $\theta_{ab}$ through an inverse noise weighted average. 

Results are shown in Fig.~\ref{fig:angular_correlations} for the data, inferred, and predicted distributions. The top panel corresponds to the \textsc{HellingsDowns-PowerLaw} dataset, while the bottom panel to \textsc{HellingsDownsMonopole-PowerLaw}. In the top panel, the inferred and predicted distributions overlap, to within expected scatter. In the bottom panel, although the distributions overlap for any given angular bin, the predicted distributions are systematically shifted downwards. This is because the inferred distributions contain a monopole, while the predicted ones are solely based on Hellings--Downs correlations.

\begin{figure}
    \centering
    \includegraphics[width=\columnwidth]{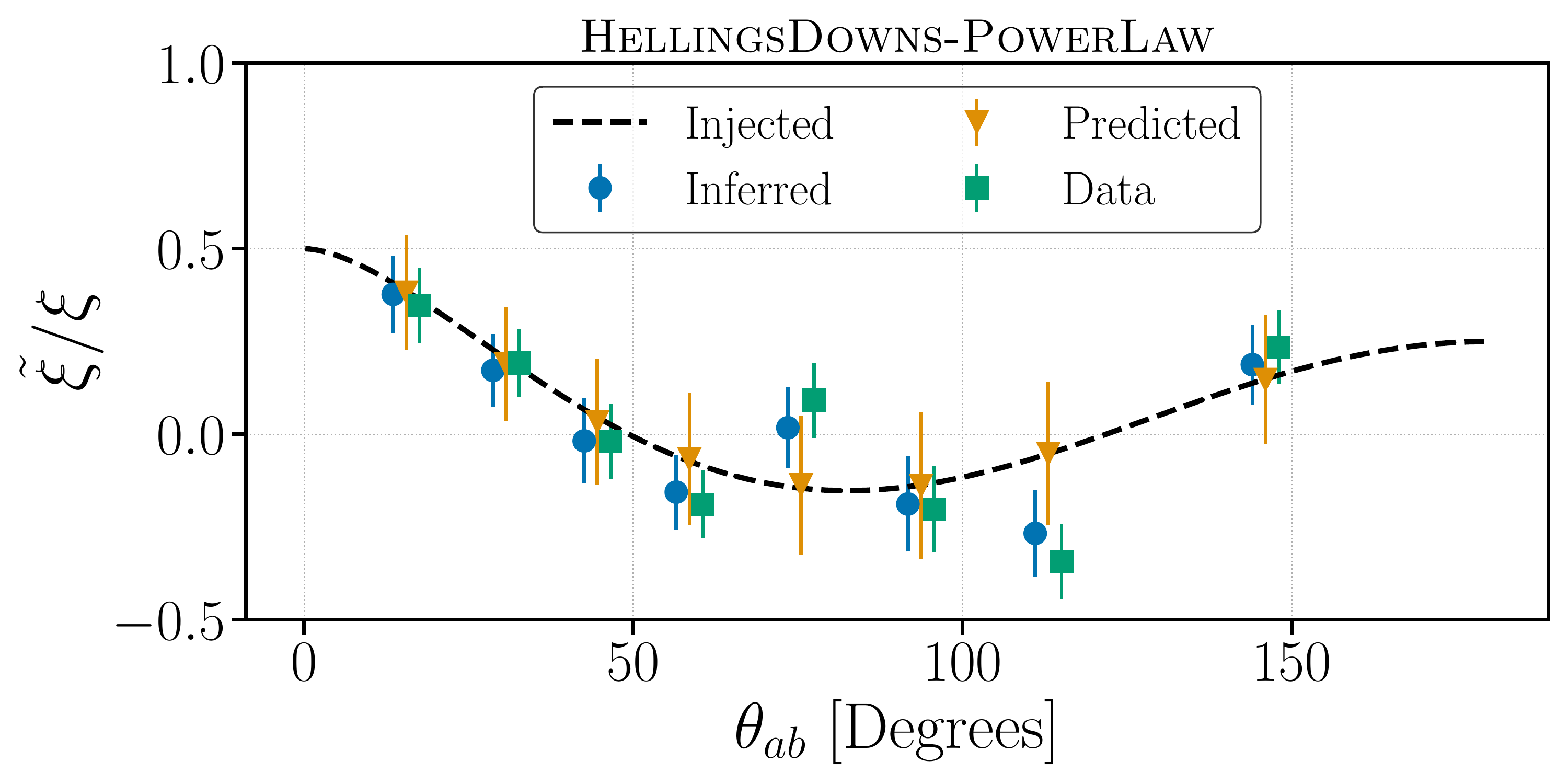}
        \includegraphics[width=\columnwidth]{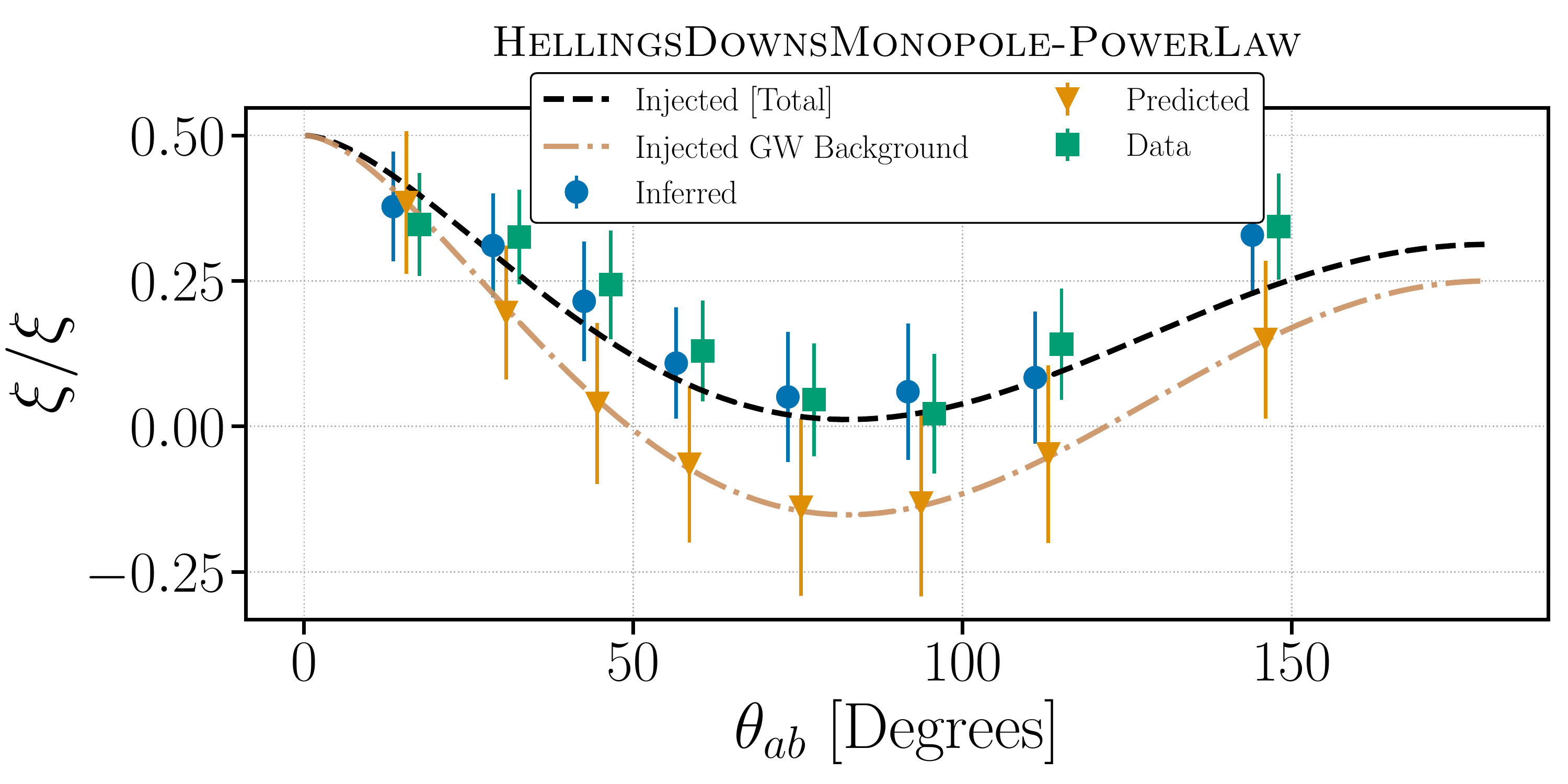}
    \caption{Spatial correlations (median and 68\% credible intervals) as a function of pulsar pair angular separation $\theta_{ab}$ for the \textsc{HellingsDowns-PowerLaw} (top) and \textsc{HellingsDownsMonopole-PowerLaw} (bottom) datasets. We show the inferred (blue) and predicted (orange) correlations as a function of pulsar angular separation. We also show the correlations as inferred from solely the data (green). The black dashed line shows the injected correlation, while the Hellings--Downs correlations are shown in orange dot-dashed in the bottom panel. In the bottom panel, the predicted correlations are systematically lower than the inferred ones.}
    \label{fig:angular_correlations}
\end{figure}

\subsubsection{Comparing spectrum and correlations mismodeling}

\begin{figure}
    \centering \includegraphics[width=\columnwidth]{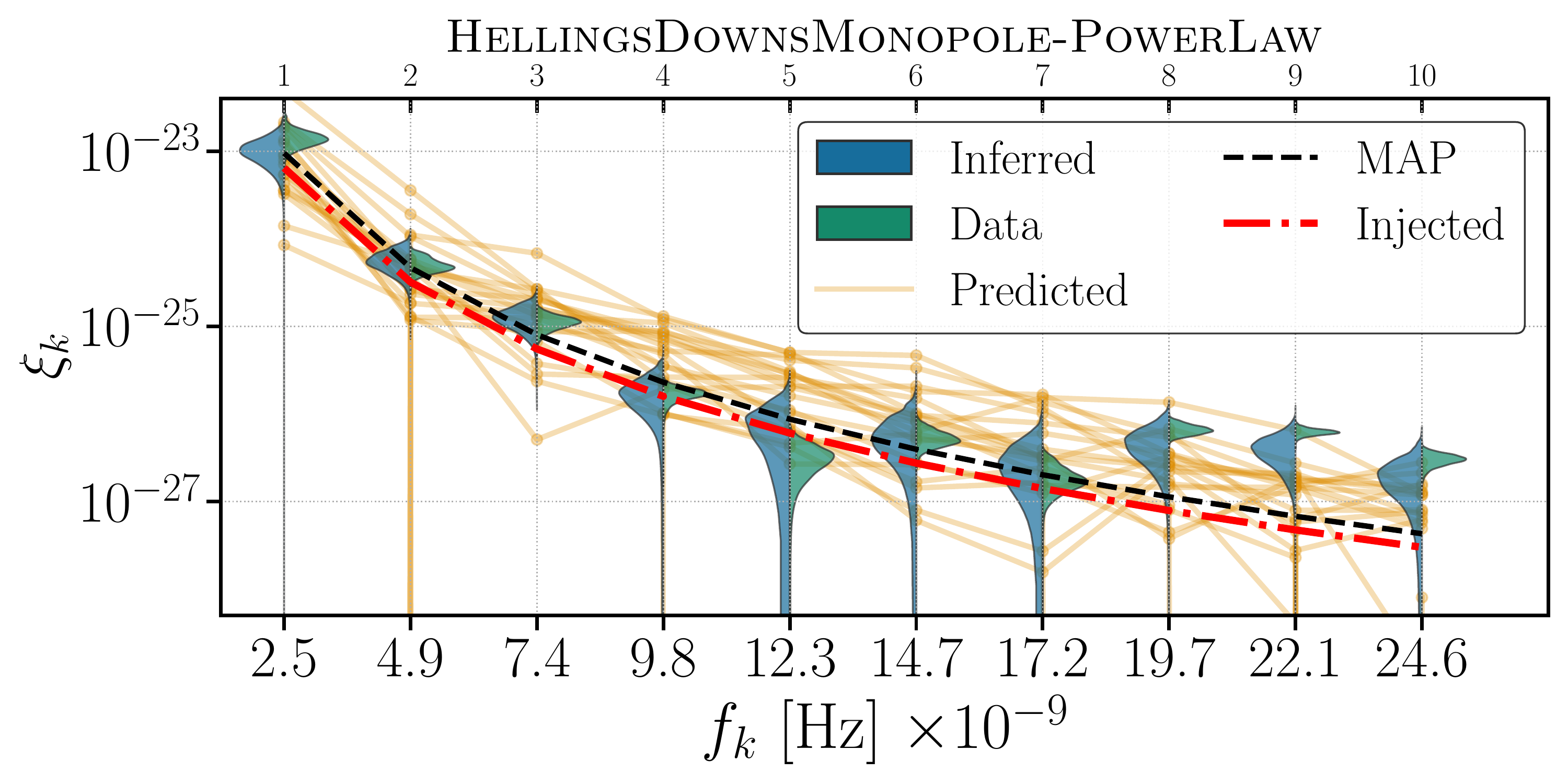}
    \includegraphics[width=\columnwidth]{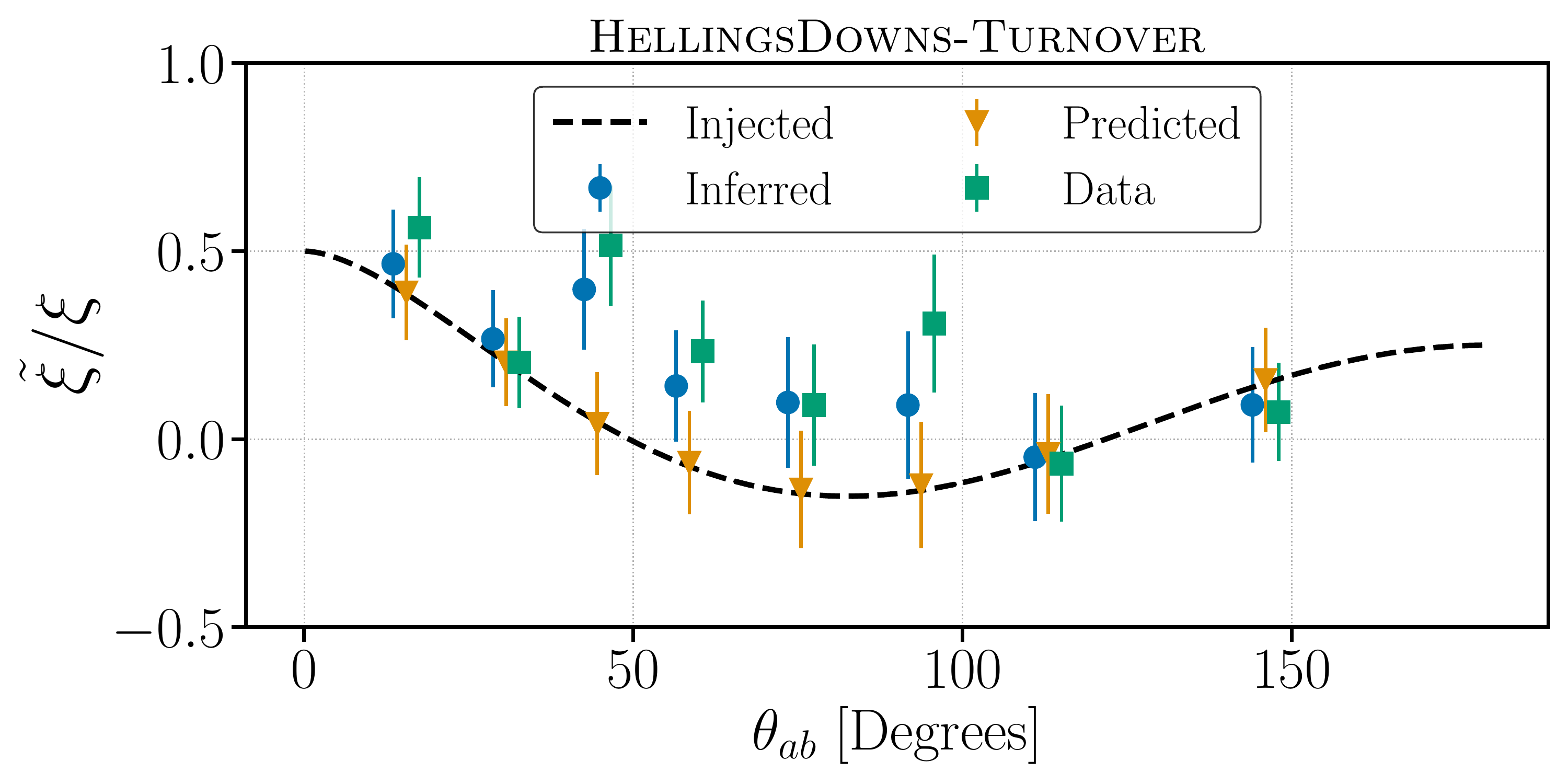}
    \caption{Total GW spectrum for the \textsc{HellingsDowns-Turnover} (bottom) and
    spatial correlations for the \textsc{HellingsDownsMonopole-PowerLaw} (top) datasets. Plotted quantities and colors are similar to Figs.~\ref{fig:total_gwb_spectra} and~\ref{fig:angular_correlations}
    A correlation mismodeling does not manifest in the spectrum comparison (top). A spectrum mismodeling has a larger effect on the characterization of the spatial correlations (bottom).}
    \label{fig:model_mismatch}
\end{figure}

The above tests demonstrate that spectral and spatial correlations mismodeling can be identified by their corresponding predictive tests. Though the spectrum and the correlation pattern of a stochastic process are separate elements of the GW model, it is not clear they are fully independent. This is because the pulsars are not uniformly distributed in the sky and the signal periods are comparable to the observation time. It is therefore possible that mismodeling in one element of the GW model appears in the test for another.
To test for such mismodeling ``leakage,'' we investigate whether using a Hellings--Downs model on the \textsc{HellingsDownsMonopole-PowerLaw} dataset can result in spectral mismodeling, and whether using a power-law model on the \textsc{HellingsDowns-Turnover} dataset can result in correlation mismodelling.

Figure~\ref{fig:model_mismatch} shows the posterior predictive comparison for the spectrum of \textsc{HellingsDownsMonopole-PowerLaw} (top) and the spatial correlations of \textsc{HellingsDowns-Turnover} (bottom). The top panel shows largely consistent inferred and predicted spectra distributions, suggesting that a mismodeling of the spatial correlations, i.e., assuming Hellings--Downs when the data also contain a monopole, does not strongly impact spectral characterization. This is likely due to the fact that spectral characterization is dominated by autocorrelations, at least for weak signals such as the ones considered here. The bottom panel shows that the predicted correlations are systematically lower than the inferred ones, which exhibit signs of a monopole, i.e. a constant upward shift. This suggests that a spectrum mismodeling can affect the inferred correlations pattern. Indeed, a misestimated GW power spectrum will affect the pulsar noise weighting in the optimal-statistic calculation, especially for informative pulsars with low intrinsic noise.

\section{\label{sec:results_timing_residuals}Predictive checks on timing residuals}

The final posterior predictive tests are based directly on the timing residuals $\tresid$. We first consider visual checks, where we use the model to predict our residuals. As in~\cite{Goncharov:2020krd}, we isolate contributions from different parts of our model, showing how they sum together to model the timing residuals. Next, we discuss \emph{leave-one-out} tests where we use data from $N_{\rm p}-1$ pulsars to predict the data of the $N_{\rm p}^{\mathrm{th}}$ pulsar. 

\subsection{\label{ssec:simulated_results:ppc} Visual data checks}

\begin{figure}
    \centering
    \includegraphics[width=0.48\textwidth]{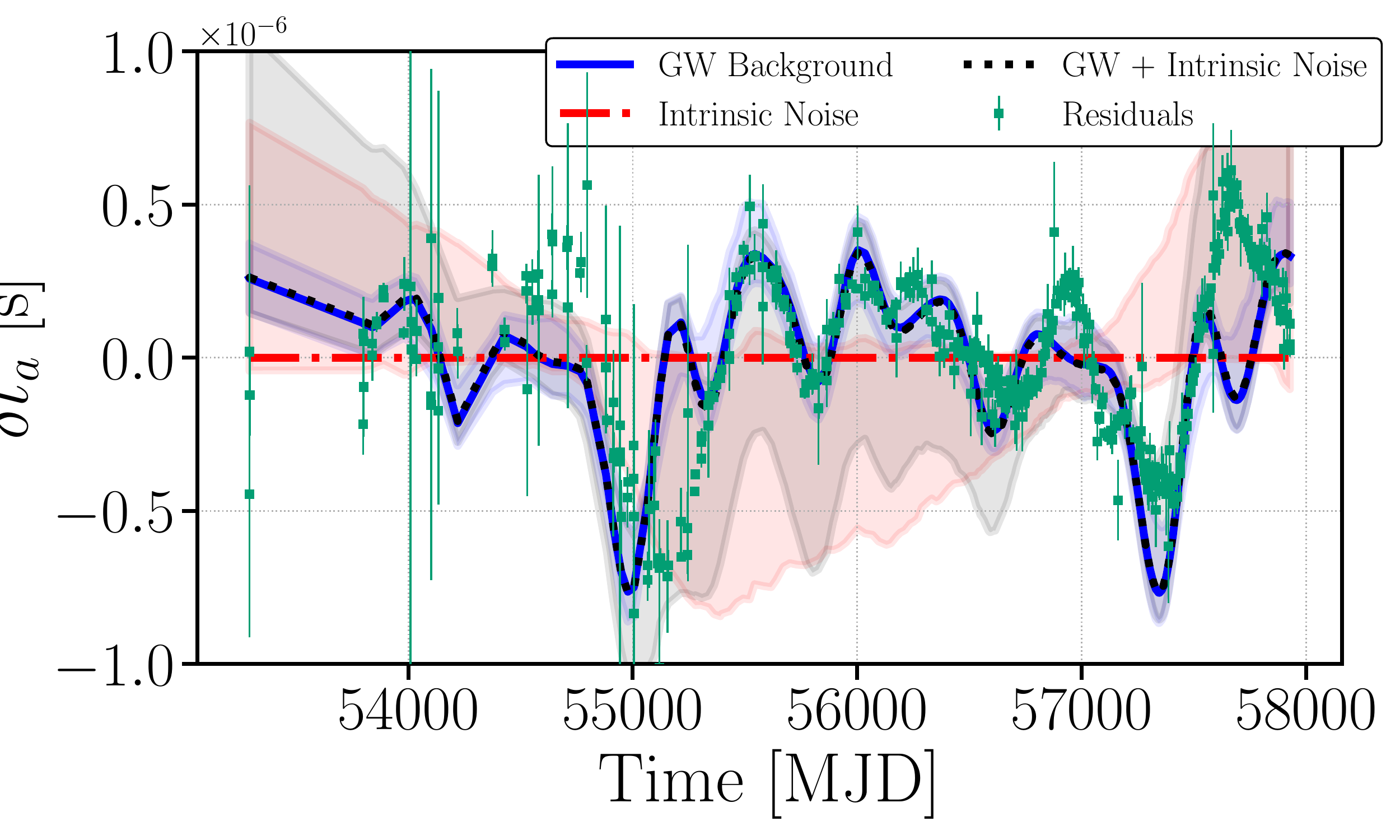}
    \includegraphics[width=0.48\textwidth]{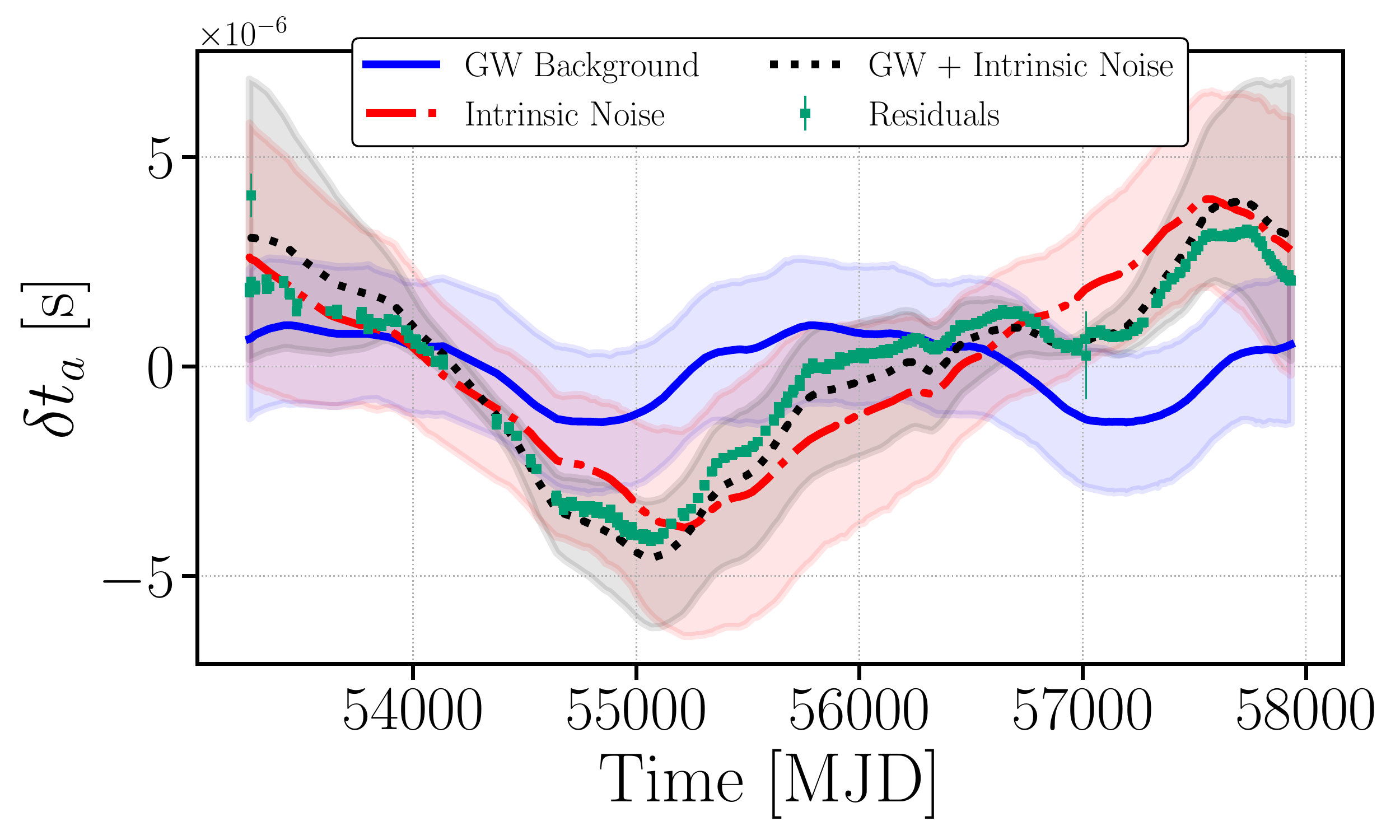}
    \caption{GW background (blue, solid), intrinsic pulsar noise (red, dashed-dotted), and total noise (black, dotted) contribution to timing residuals for J1909$-$3744 (top) and B1937+21 (bottom), compared to the simulated residuals (green). The shaded regions indicate 90\% credible intervals and the lines indicate the median. The residuals were simulated using the \textsc{HellingsDowns-Turnover} model. For J1909$-$3744, the residuals are dominated by the GW, while for B1937+21 the intrinsic noise dominates. In both cases, the total noise posterior tracks the residuals closely. We do not plot the timing model corrections for clarity, as they are small in this case.}
    \label{fig:reconstruction_and_prediction}
\end{figure}

We use the Gaussian process coefficients from Sec.~\ref{sec:results-parameters} to reconstruct expected residuals for each pulsar. We draw $\bvec_{a}^s \sim p(\bvec_{a} | \bm\Lambda^{s}, \tresid)$, and use these to reconstruct predicted timing residuals in pulsar $a$, $\tresid_{a}^s = \tmat_a\bvec_{a}^s$. This procedure allow us to separate contributions to the residuals from the GW background, the intrinsic pulsar noise, and from timing-model fluctuations.

Figure~\ref{fig:reconstruction_and_prediction} plots the simulated timing residuals and the separate contributions from intrinsic pulsar noise, GW background, and the sum of the two for J1909$-$3744 (top, low intrinsic noise) and B1937+21 (bottom, high intrinsic noise) for the \textsc{HellingsDowns-Turnover} dataset. These reconstructions include frequencies $f_i > 3 / T$, because the two lowest frequencies are degenerate with the frequency and spin down parameters in the timing model. In the J1909$-$3744 case (top) the median estimate of the intrinsic noise at each time is near zero, although there is a spread in potential values. Meanwhile, the GW background and total noise (GW plus intrinsic) track the residuals more closely. In the B1937+21 case (bottom) the residuals are dominated by intrinsic noise, while the GW background contribution is smaller.

We do not show the contribution from timing-model corrections as it is small in this case. However, their posterior is estimated and could be compared to the fiducial values used to create the original timing residuals. This could serve as a useful cross-check, especially for individual pulsars that are difficult to model. 

\subsection{\label{ssec:model_comparison}Leave-one-out analysis: Hellings--Downs vs common noise model comparison}
We construct predicted data distributions for each pulsar under different assumptions for the correlation pattern, and specifically assuming either Hellings--Downs correlations or an uncorrelated common process. 
Evaluating these distributions on the actual observed data, we introduce a \emph{pseudo Bayes factor}~\cite{geiser_eddy_pbf} for the presence of Hellings--Downs correlations. We compare the pseudo Bayes factor to null distributions obtained from simulated data and show how they can be used to establish the presence of Hellings--Downs correlations, and equivalently the detection of a GW background.

In contrast to the parameter predictive tests of Sec.~\ref{sec:results-parameters}, here we perform per-pulsar tests conditioned on the data of the \emph{other} pulsars. This distinction is driven by two main reasons. Firstly, the tests of Sec.~\ref{sec:results-parameters} focus on GW model parameters, inference of which is informed by more than one pulsar. For example, the GW Gaussian process coefficients in one pulsar are informed by the other pulsars through Hellings--Downs correlations. There is therefore no clear sense in which GW parameters ``belong" to one pulsar.
Secondly, typically a small number of pulsars dominates the constraints. Therefore in-sample and out-of-sample data predictions can be quite distinct.

We begin by selecting a pulsar $a$ to leave out. Quantities with a subscript of $a$ correspond to this pulsar, while a subscript of $-a$ denotes the set of all the other pulsars in the array. We also explicitly break up all quantities into GW, pulsar $a$, and all other pulsars ($-a$): $\bm \epsilon=[\bm \epsilon_{a},\bm \epsilon_{-a}]$, $\bm \Lambda=[\bm\Lambda_{\rm gw},\bm \Lambda_{a},\bm \Lambda_{-a}]$, $\avec=[\avec_{{\rm gw},a},\avec_{{\rm gw},-a},\avec_{a},\avec_{-a}]$. This split is motivated by the fact that $\tresid_{-a}$ offers no information about the intrinsic parameters of pulsar $a$, for example $p(\bm\Lambda | \tresid_{-a}) = p(\bm\Lambda_{\rm gw},\bm\Lambda_{-a} | \tresid_{-a})p(\bm\Lambda_a)$.

The likelihood of residuals $\tresid_{a}$ in pulsar $a$ given the residuals $\tresid_{-a}$ in all other pulsars is
\begin{align}
\label{eq:pta_loo_posterior_hd}
    p(\tresid_a | \tresid_{-a}) & = \int \textrm{d}\bm\Lambda \, \textrm{d}\bm\epsilon \,\textrm{d}\avec \,\, p(\tresid_a | \bm\Lambda, \bm\epsilon, \avec)p(\bm\Lambda, \bm\epsilon, \avec|\tresid_{-a})\,.
\end{align}
After a long derivation laid out in App.~\ref{app:PPL derivation} we find 
\begin{widetext}
\begin{align}
   p_{\textrm{HD}}(\tresid_a | \tresid_{-a}) 
    & \approx \frac{1}{N_s}\sum_{s} \int \textrm{d}\bm\Lambda_{a}\textrm{d}\avec_{{\rm gw},a} p(\tresid_a | \bm\Lambda_{a},\avec_{\textrm{gw}, a}) p(\avec_{\textrm{gw}, a} | \bm\Lambda_{\rm gw}^{s}, \bm\Lambda_{-a}^{s}, \tresid_{-a})p(\bm\Lambda_{a})\,, \label{eq:pta_loo_posterior_hd_monte_carlo_mt}\\
    p_{\textrm{CN}}(\tresid_a | \tresid_{-a}) &\approx \frac{1}{N_s}\sum_{s}\int \textrm{d}\bm\Lambda_{a}p(\tresid_a | \bm\Lambda_{a},\bm\Lambda_{\rm gw}^{s})p(\bm\Lambda_{a})\,.
    \label{eq:crn_monte_carlo_ppl_mt}
\end{align}
\end{widetext}
where the ``HD'' subscript signifies that we have assumed Hellings--Downs correlations and ``CN'' subscript signifies that we ignore the Hellings--Downs correlations and assume that the pulsars are only subject to an uncorrelated common process. Equations~\eqref{eq:pta_loo_posterior_hd_monte_carlo_mt} and~\eqref{eq:crn_monte_carlo_ppl_mt} are evaluated over $N_s$ draws from the hyperparameter posterior
\begin{align}
    \bm\Lambda^{s}_{\rm gw}, \bm\Lambda^{s}_{-a} &\sim p(\bm\Lambda_{\rm gw}, \bm\Lambda^{s}_{-a} | \tresid_{-a})\,,
\end{align}
from the analysis of Sec.~\ref{sec:datasets}. The integral over $\textrm{d}\avec_{{\rm gw},a}$ is performed analytically as it involves a product of Gaussian distributions, while the one over $\textrm{d}\bm \Lambda_{a}$ is performed numerically.

Comparing Eqs.~\eqref{eq:pta_loo_posterior_hd_monte_carlo_mt} and~\eqref{eq:crn_monte_carlo_ppl_mt} can provide an estimate of how much each pulsar supports the presence of Hellings--Downs correlations. We introduce the ``pseudo Bayes factor'' (PBF)~\cite{geiser_eddy_pbf} between Hellings--Downs and common noise in pulsar $a$ as
\begin{equation}
    \textrm{PBF}^{\textrm{HD}}_{\textrm{CN},a} \equiv \frac{p_{\textrm{HD}}(\tresid_a | \tresid_{-a}) }{p_{\textrm{CN}}(\tresid_a | \tresid_{-a}) }\,,
\end{equation}
where the numerator and denominator are defined in Eqs.~\eqref{eq:pta_loo_posterior_hd_monte_carlo_mt} and~\eqref{eq:crn_monte_carlo_ppl_mt} respectively, and are \textit{posterior predictive likelihoods} that are calculated on the observed $\tresid_a$. The total pseudo Bayes factor is then the product over all pulsars
\begin{equation}
\textrm{PBF}^{\textrm{HD}}_{\textrm{CN}} = \prod_{a=1}^{N_{\textrm{p}}}\textrm{PBF}^{\textrm{HD}}_{\textrm{CN},a}\,.
\end{equation}
The pseudo Bayes factor shares some similarities with the traditional Bayes factor (i.e., the marginal likelihood ratio), but there are also important differences. 
First, both traditional and pseudo Bayes factors are a ratio of likelihoods. Second, unlike traditional Bayes factors, the pseudo Bayes factor is insensitive to the existence of parameter space regions of little likelihood support, which reduce Bayes factors by the so-called Occam factors. In that sense, the pseudo Bayes factor does not suffer from interpretation issues related to the extent of parameter priors or the presence of improper priors~\cite{Hazboun:2020kzd,Isi:2022cii,Zic:2022sxd}. Third, by definition $\textrm{PBF}^{\textrm{HD}}_{\textrm{CN},a}$ is a measure of how well the model predicts new data. This means that it can be estimated on a per-pulsar basis, thereby assessing which pulsar is more consistent with each model, and identifying outliers. Specifically, $\textrm{PBF}^{\textrm{HD}}_{\textrm{CN},a}$ tests whether certain pulsars are poorly understood compared to others, potentially signaling issues with their intrinsic noise modeling.

The pseudo Bayes factor, however, does suffer from \emph{calibration} issues just as the traditional Bayes factor. That is, how are we to interpret its value in terms of statistical confidence?
Rather than relying on arbitrary classifications schemes~\cite{jeffreys1998theory,kassraftery1995}, a common procedure to interpret Bayes factors involves using a large set of simulations to estimate a false-alarm probability for the measured value~\cite{Veitch:2008wd,2012PhRvD..86h2001V,Cornish:2015ikx,Taylor:2016gpq,Littenberg:2015kpb,Isi:2018vst}.\footnote{Another recommendation is to compare $ \textrm{PBF}^{\textrm{HD}}_{\textrm{CN}}$ to the variance of $ \textrm{PBF}^{\textrm{HD}}_{\textrm{CN},a}$ over the pulsars \cite{2015arXiv150704544V}; we leave this to future work.} 

\begin{figure}
    \centering
    \includegraphics[width=0.49\textwidth]{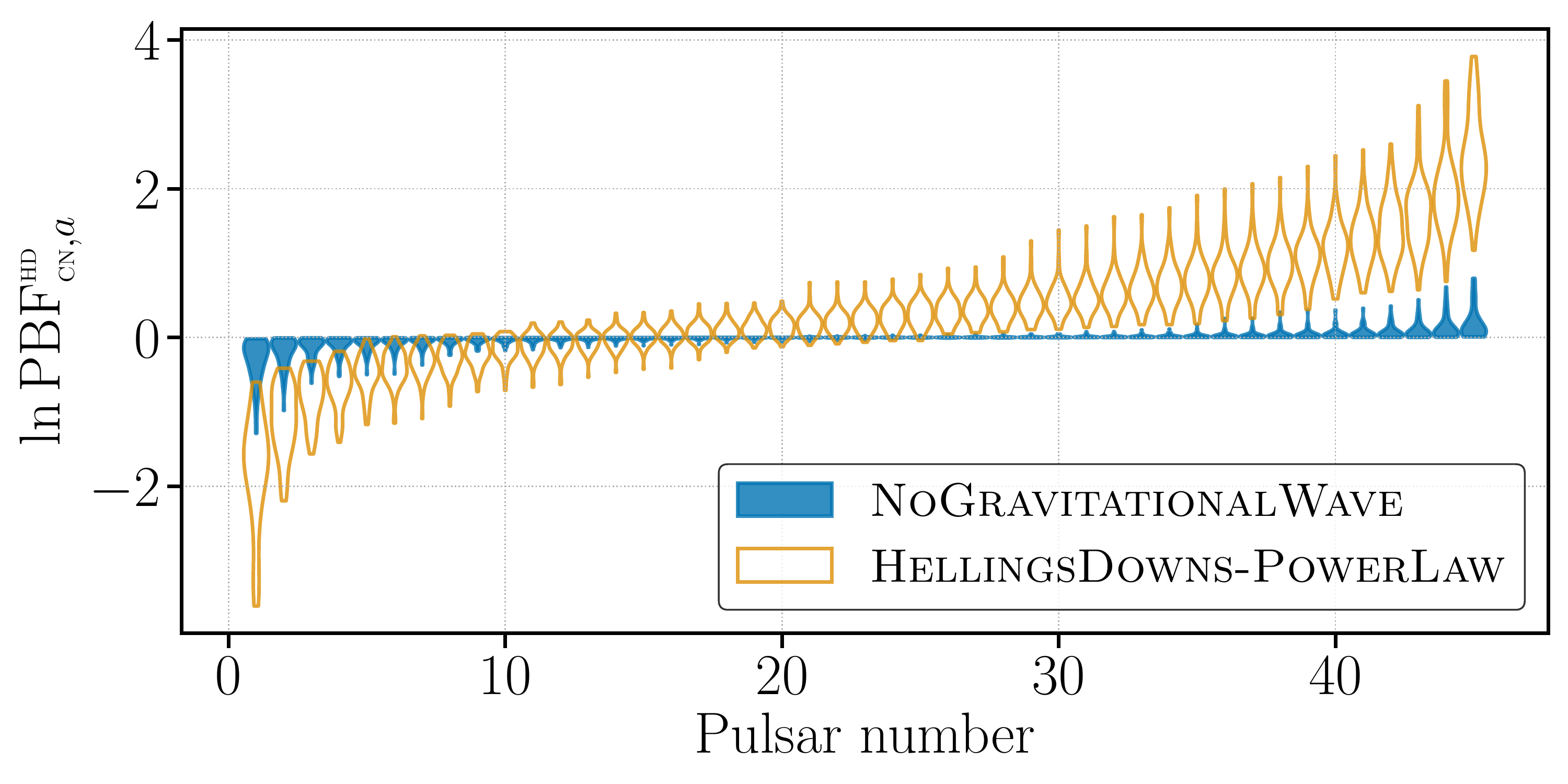}
    \caption{Pseudo Bayes factor comparing the Hellings--Downs and common noise models for each pulsar. We consider $59$ realizations of simulated \textsc{HellingsDowns-PowerLaw} (orange) and $45$ \textsc{NoGravitationalWave} (blue) datasets and show the distribution of obtained pseudo Bayes factors in violins. Pulsars are ordered from lowest to higher value of the pseudo Bayes factor. Most pulsars support the presence of Hellings--Downs correlations when a signal is present, though a minority displays the opposite behavior. All pulsars have uninformative pseudo Bayes factors when no signal is injected. 
    }
    \label{fig:pbf_dropout_factor}
\end{figure}

Figure~\ref{fig:pbf_dropout_factor} shows the (natural logarithm of the) pseudo Bayes factors for individual pulsars ordered from lowest to highest. We produce $59$ simulated datasets using \textsc{HellingsDowns-PowerLaw}, and $45$ using \textsc{NoGravitationalWave}. The following result should be interpreted only as a demonstration of our method as the simulated GW background amplitude of $\log_{10} A_{\textrm{gw}}=-14$ is higher than the one inferred from real data by a factor of $\sim5$~\cite{NANOGrav:2020bcs}. Such a high value was chosen so that we have a detectable signal in $12$ years of simulated data and thus we can meaningfully test the proposed methods. 

For each simulated dataset, we compute $\ln\textrm{PBF}^{\textrm{HD}}_{\textrm{CN},a}$ for each pulsar $a$, we sort the pulsars from the smallest to the largest value, and we plot the distribution over data realizations.\footnote{This procedure means that the pulsar order is different for each simulated dataset. Therefore the x-axis of Fig.~\ref{fig:pbf_dropout_factor} is not a \emph{specific} pulsar, but instead the $n^{th}$ pulsar as ranked by its pseudo Bayes factor in each dataset.}
In the \textsc{HellingsDowns-PowerLaw} case, we regularly find $\sim 20$ pulsars with positive $\ln\textrm{PBF}^{\textrm{HD}}_{\textrm{CN},a}$. This means that data from the other pulsars can predict the observed data in pulsar $a$ \emph{better} if Hellings--Downs correlations are present. The test is uninformative for $\sim 10$ pulsars with $\ln\textrm{PBF}^{\textrm{HD}}_{\textrm{CN},a}\sim0$, while a similar number of pulsars has $\ln\textrm{PBF}^{\textrm{HD}}_{\textrm{CN},a}<0$. The latter means that these pulsars support no Hellings--Downs correlations even if these exist in the data. Such behavior is also encountered in the ``drop-out factors''  calculated by sampling an indicator variable that switches between the common process and no common signal hypotheses for each individual pulsar~\cite{NANOGrav:2020bcs}. \emph{Some} negative $\ln\textrm{PBF}^{\textrm{HD}}_{\textrm{CN},a}$ are therefore to be expected even in simulated data and they are not immediately an indication of mismodeling.\footnote{In fact down-selecting pulsars based on arbitrary metrics can lead to biased estimates~\cite{Johnson:2022uxn}.} In the \textsc{NoGravitationalWave} case, all pulsars have $\ln\textrm{PBF}^{\textrm{HD}}_{\textrm{CN},a}\sim0$, suggesting no preference either way. This is to be expected as no signal is present, so there should be no information about its correlation pattern. 

Even though individual pulsars can have $\ln\textrm{PBF}^{\textrm{HD}}_{\textrm{CN},a}<0$, the total pseudo Bayes factor is in favor of Hellings--Downs correlations for the majority of the simulated datasets with a signal. Figure~\ref{fig:pbf_pbf_distributions} shows distributions of $\ln\textrm{PBF}^{\textrm{HD}}_{\textrm{CN}}$ over $59$ data realizations for \textsc{HellingsDowns-PowerLaw} (top) and $45$ for \textsc{NoGravitationalWave} (bottom). In the top panel, we find $\ln\textrm{PBF}^{\textrm{HD}}_{\textrm{CN}}>0$ for $92\%$ of the realizations, with most datasets resulting in a strong preference for Hellings--Downs correlations and $\ln\textrm{PBF}^{\textrm{HD}}_{\textrm{CN}}\sim 10-20$. However, as discussed above, the absolute scale of the pseudo Bayes factor has no definite statistical interpretation, and results should instead be calibrated to simulations. The bottom panel shows the null distribution of $\ln\textrm{PBF}^{\textrm{HD}}_{\textrm{CN}}$. All datasets have $\ln\textrm{PBF}^{\textrm{HD}}_{\textrm{CN}}<2$ and $61\%$ of them have $\ln\textrm{PBF}^{\textrm{HD}}_{\textrm{CN}}<0$. Given this null, Hellings--Downs correlations would have been detected in $89\%$ of the \textsc{HellingsDowns-PowerLaw} simulations with a significance of $> 2\sigma$. With $59$ background simulations the significance estimate is limited to $\sim 1/59 \sim 2\sigma$.

Figure~\ref{fig:pbf_pbf_distributions} shows also the distributions of traditional Bayes factors between the Hellings--Downs and common noise hypotheses for the same simulations computed via likelihood reweighting~\cite{Hourihane:2022ner}. On average the \textsc{HellingsDowns-PowerLaw} dataset results in larger pseudo Bayes factors than traditional Bayes factors, while the trend is reversed for the \textsc{NoGravitationalWave} datasets. However, due to the high GW signal amplitude we still find that 90\% of the simulated datasets in the top panel have detectable Hellings--Downs correlations at $>2\sigma$ significance when using the traditional Bayes factor as a detection statistic. These results suggest that pseudo and traditional Bayes factors can act as complementary model-checking tools. We leave the determination of their relative sensitivity as detection statistics to future work, since this demonstration is based on only $45$ simulations and a loud injected GWB.

\begin{figure}
    \centering
    \includegraphics[width=0.49\textwidth]{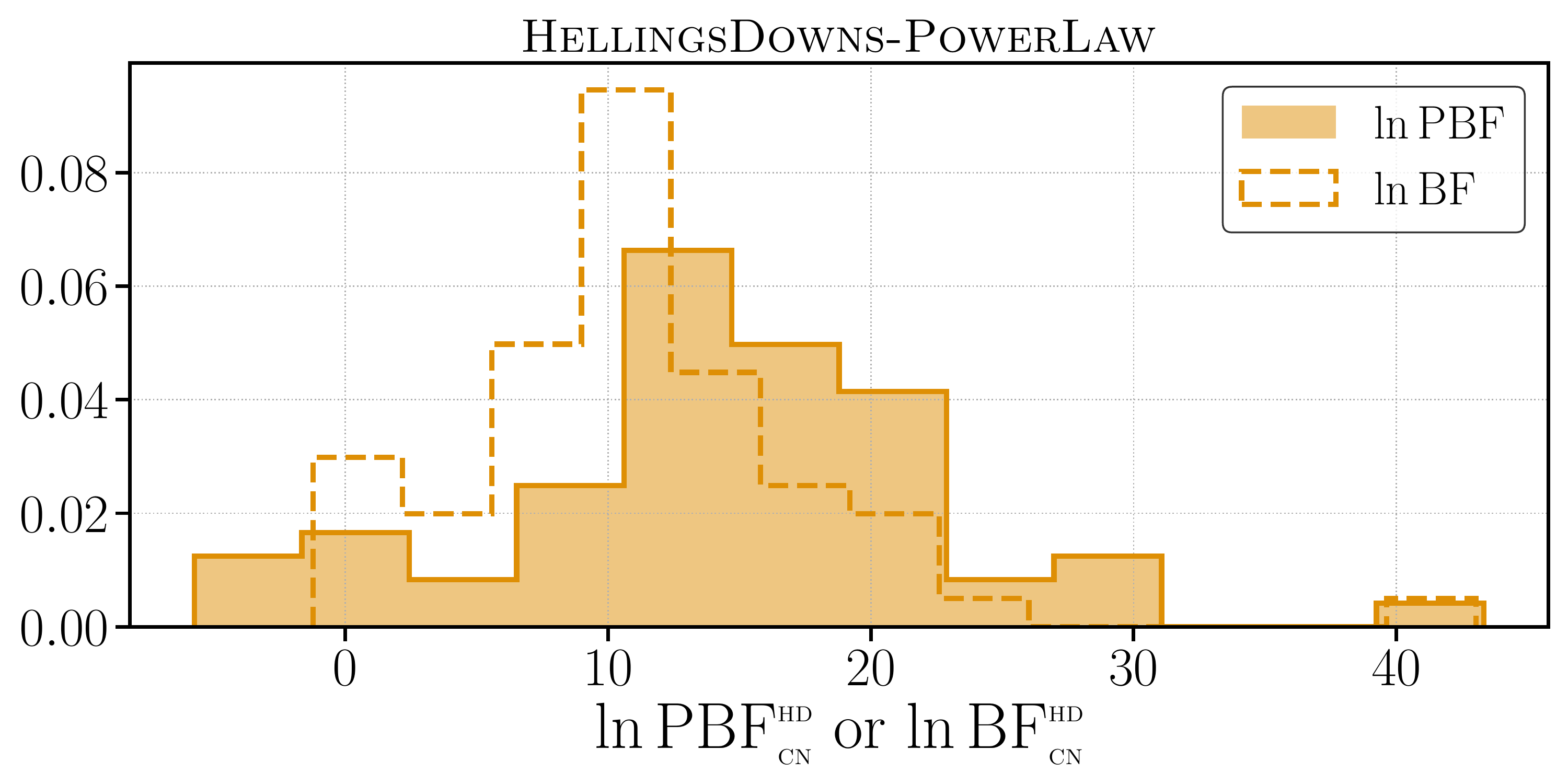}
    \includegraphics[width=0.49\textwidth]{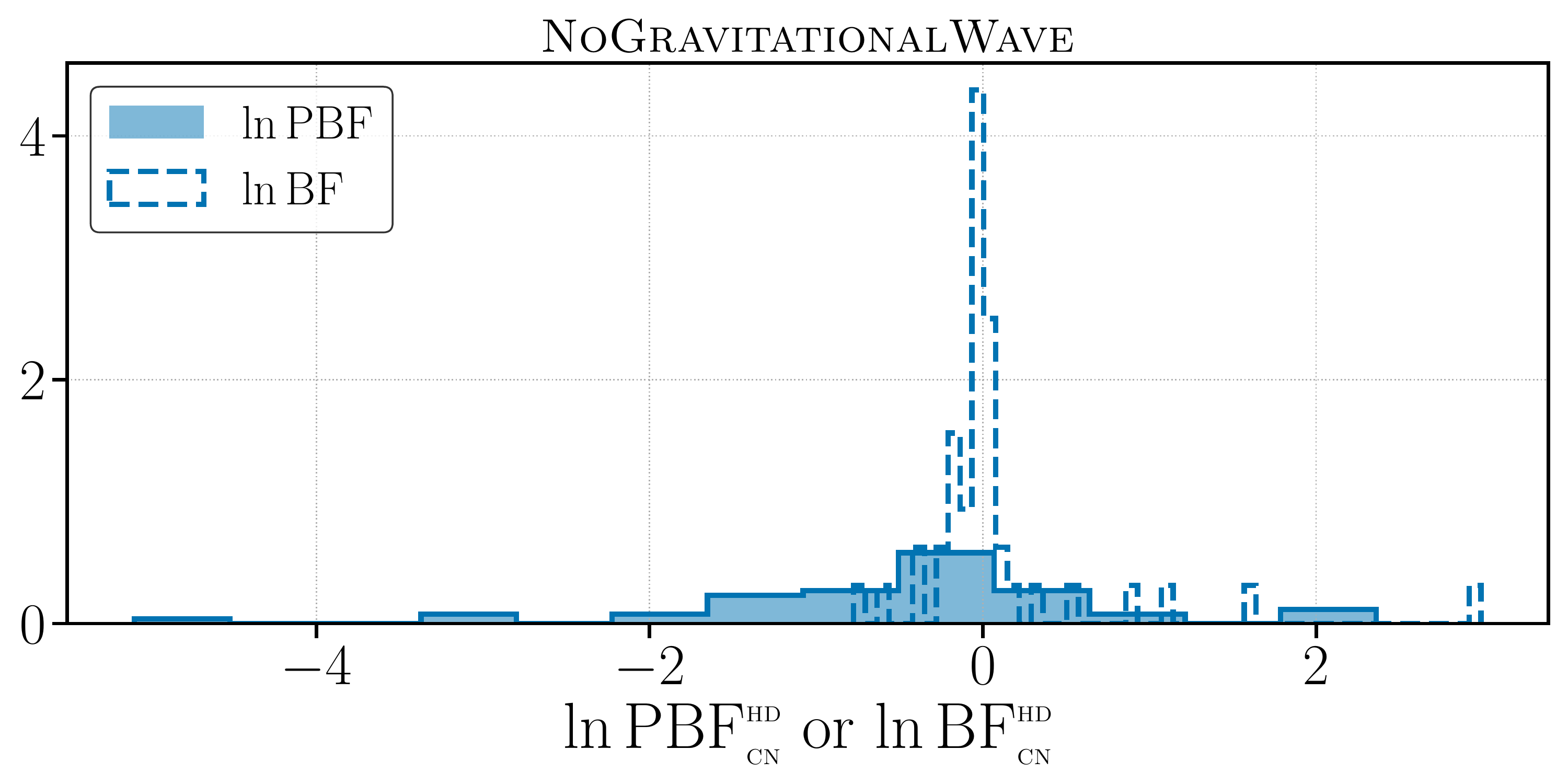}
    
    \caption{Distribution of total pseudo Bayes factors (solid histograms) and traditional Bayes Factors (dashed histograms) for repeated simulations with the \textsc{HellingsDowns-PowerLaw} (top) and \textsc{NoGravitationalWave} (bottom) datasets comparing the Hellings--Downs and common noise hypotheses. Note the different $x$-axis scales on the two panels. Using the results of the bottom panel as a null distribution, $89$\% of the simulated datasets in the top panel have detectable Hellings--Downs correlations at $>2\sigma$ significance.
    }
    \label{fig:pbf_pbf_distributions}
\end{figure}

\section{\label{sec:discussion_and_conclusion}Discussion and conclusions}

PTA analyses assume that a GW background results in arrival time residuals that are subject to a common power-law process among pulsars and Hellings--Downs spatial correlations between them. While the correlation pattern is robust under a tensorial GW background, systematic errors can induce further monopolar or dipolar correlations~\cite{Tinto:2018bae, Tiburzi:2015kqa,Goncharov:2020krd, NANOGrav:2020tig_bayesephem, Caballero:2018lvc, Roebber:2019gha}. Moreover, the GW spectral shape is subject to astrophysical, statistical, and even cosmological uncertainties~\cite{Middleton:2020asl,Becsy:2022pnr,Taylor:2016gpq}. Here we propose to test these assumptions using \emph{posterior predictive checks} that assess how well predicted data based on the inferred model parameters match the observed data. Predictive tests based on different quantities allow us to assess different aspects of the model or pulsars in the array separately and thus can offer insights about model extensions if a discrepancy is identified.

We propose and study two types of tests. The first type concerns the Gaussian-process coefficients of the GW and intrinsic-noise stochastic processes. Comparing predicted and inferred coefficients on simulated datasets, we can identify frequency bins where the power-law model under- or over-predicts the observed power. Moreover, by comparing the inferred and predicted spatial correlations we can assess the presence of non-Hellings--Downs correlations. 
The second type of test concerns the timing residuals themselves, and specifically the likelihood of the observed data in a select pulsar given all other pulsars. We compute the \emph{pseudo Bayes factor} as the ratio of these likelihoods under the Hellings--Downs and the uncorrelated common process hypotheses. We show that among all the pulsars in the array it is expected for a handful to show preference \emph{against} Hellings--Downs correlations.
However, the total pseudo Bayes factor over the entire array can be used as a detection statistic to establish the presence of Hellings--Downs correlations.
 
Our study adds to existing efforts that explore extensions of PTA analyses. 
A common extension to the power-law spectrum (and one of our simulated datasets) is the truncated power-law that arises when astrophysical hardening mechanisms accelerate the inspiral of the black hole binaries that source the GW background~\cite{Sampson:2015ada}.
A different kind of broken power-law flattens the spectrum at high frequencies~\cite{NANOGrav:2020bcs}. Such flattening is interpreted as being caused by modeling systematics related to the intrinsic pulsar noise, and it is used to limit the number of frequency bins analyzed~\cite{NANOGrav:2020bcs}.
Doing away with a parametric model, ``free spectral" analyses instead allow for independent amplitudes at each frequency bin~\cite{NANOGrav:2020bcs}.
Beyond the details of the spectral shape, a GW background has a unique spectrum, even though the exact realization will differ between pulsars. A test of this assumption involves allowing for some scatter in the GW amplitude inferred from each pulsar, whose probable origin would be mismodeling~\cite{Goncharov:2022ktc}. Applying the test to PPTA data, Ref.~\cite{Goncharov:2022ktc} found no evidence for such a scatter. 

Moving on to spatial correlations, proposed checks include reconstructing the correlations as interpolated functions, sums of Legendre polynomials~\cite{Gair:2014rwa}, or perturbed Hellings--Downs patterns~\cite{Taylor:2012wv}. These tests proceed with the observed data alone and compare the reconstructed generic correlation pattern with the expected Hellings--Downs pattern. A related test replaces or augments the Hellings--Downs correlations with non-tensorial correlations expected for certain theories of gravity beyond General Relativity~\cite{Chen:2021wdo,NANOGrav:2021ini}. 

The tests proposed in this study offer complementary ways to assess PTA models. We expect such tests to become increasingly important as PTA datasets expand in sensitivity, and move toward detection of the GW background. Furthermore, our tests can be used to assess consistency between different PTA datasets. For example, we could use NANOGrav data to predict PPTA data and then compare to the actual observed PPTA data. Such tests would generalize the comparisons performed in~\cite{Antoniadis:2022pcn} and help establish consistency between datasets, thus strengthening astrophysical conclusions.

\acknowledgements

We thank Will Farr for discussions on posterior predictive checks in the LIGO context.
Our analyses make use of \texttt{Enterprise}~\cite{enterprise, enterpriseExtensions},
\texttt{scipy}~\cite{2020SciPy-NMeth}, \texttt{matplotlib}~\cite{Hunter:2007}, \texttt{numpy}~\cite{harris2020array}, \texttt{pandas}~\cite{reback2020pandas}, and \texttt{seaborn}~\cite{Waskom2021}.
This work used the Extreme Science and Engineering Discovery Environment (XSEDE), supported by NSF award ACI-1548562, and specifically the Bridges-2 system at the Pittsburgh Supercomputing Center, supported by NSF award ACI-1928147.
PMM, MV, and KC were supported by the NANOGrav Physics Frontiers Center, National Science Foundation (NSF), Grant No. 2020265.
Part of this research was carried out at the Jet Propulsion Laboratory, California Institute of Technology, under a contract with the National Aeronautics and Space Administration (80NM0018D0004).
Copyright 2023. All rights reserved.

\appendix
\section{Detailed derivation of the posterior predictive likelihood for single-pulsar data}
\label{app:PPL derivation}

The starting point of the derivation is the likelihood of the residuals $\tresid_{a}$ in pulsar $a$ given the residuals $\tresid_{-a}$ in all other pulsars, reproduced here from Eq.~\eqref{eq:pta_loo_posterior_hd}: 
\begin{align}
\label{eq:pta_loo_posterior_hd_app}
    p(\tresid_a | \tresid_{-a}) & = \int \textrm{d}\bm\Lambda \, \textrm{d}\bm\epsilon \,\textrm{d}\avec \,\, p(\tresid_a | \bm\Lambda, \bm\epsilon, \avec)p(\bm\Lambda, \bm\epsilon, \avec|\tresid_{-a})\,.
\end{align}
The first term in the integrand of Eq.~\eqref{eq:pta_loo_posterior_hd_app} reduces to 
\begin{equation}
p(\tresid_a | \bm\Lambda, \bm\epsilon, \avec) = p(\tresid_a | \bm\epsilon_{a}, \avec_{{\rm gw},a},\avec_{a})\label{eq:predictivelikelihoodfirstterm}\,,
\end{equation}
as the data of pulsar $a$ depend on the parameters of this pulsar only, as given by Eq.~\eqref{eq:pre_fit_resids_likelihood_gp_coefficients}. The second term in the integrand of Eq.~\eqref{eq:pta_loo_posterior_hd_app} is
\begin{widetext}
\begin{align}
p(\bm\Lambda, \bm\epsilon, \avec|\tresid_{-a})&=p(\bm\Lambda_{a},\bm\epsilon_{a},\avec_{a})  p(\bm\Lambda_{\rm gw},\bm\Lambda_{-a},\bm\epsilon_{-a},\avec_{-a},\avec_{{\rm gw},a},\avec_{{\rm gw},-a}|\tresid_{-a})\label{eq:predictivelikelihoodsecondterm}\,,
\end{align}
\end{widetext}
where the first term includes all properties of pulsar $a$ that do not depend on the data of the other pulsars. The only property of pulsar $a$ that remains in the second term are the GW Gaussian process coefficients $\avec_{{\rm gw},a}$, since those are informed by $\tresid_{-a}$ through the Hellings--Downs correlations. Returning to the full predictive likelihood in Eq.~\eqref{eq:pta_loo_posterior_hd_app}, the integrals over $\avec_{{\rm gw},-a},\bm\epsilon_{-a},\avec_{-a}$ are now trivial. Performing those and substituting Eqs.~\eqref{eq:predictivelikelihoodfirstterm} and~\eqref{eq:predictivelikelihoodsecondterm} in Eq.~\eqref{eq:pta_loo_posterior_hd_app} we get
\begin{widetext}
    \begin{align}
        p_{{\rm HD}}(\tresid_a | \tresid_{-a}) & = \int \textrm{d}\bm\Lambda_{\rm gw} \textrm{d}\bm \Lambda_{a} \, \textrm{d}\bm \Lambda_{-a} \, \textrm{d}\bm\epsilon_{a} \,\textrm{d}\avec_{{\rm gw},a} \textrm{d} \avec_{a} \,\, p(\tresid_a | \bm\epsilon_{a}, \avec_{{\rm gw},a},\avec_{a})p(\bm\Lambda_{a},\bm\epsilon_{a},\avec_{a})  p(\bm\Lambda_{\rm gw},\bm\Lambda_{\rm -a},\avec_{{\rm gw},a}|\tresid_{-a})\nonumber
        \\
        &= \int \textrm{d}\bm\Lambda_{\rm gw} \textrm{d}\bm \Lambda_{a} \,
        \textrm{d}\bm \Lambda_{-a} \, \textrm{d}\bm\epsilon_{a} \,\textrm{d}\avec_{{\rm gw},a} \textrm{d} \avec_{a} \,\, p(\tresid_a | \bm\epsilon_{a}, \avec_{{\rm gw},a},\avec_{a})p(\bm\Lambda_{a})p(\bm\epsilon_{a},\avec_{a}|\bm\Lambda_{a})  p(\bm\Lambda_{\rm gw},\bm\Lambda_{\rm -a},\avec_{{\rm gw},a}|\tresid_{-a})\nonumber
        \\
        &= \int \textrm{d}\bm\Lambda_{\rm gw} \textrm{d}\bm \Lambda_{a}  \, \textrm{d}\bm \Lambda_{-a}  \, \textrm{d}\avec_{{\rm gw},a} \,\, p(\tresid_a | \avec_{{\rm gw},a},\bm\Lambda_{a}) p(\bm\Lambda_{a})p(\bm\Lambda_{\rm gw},\bm\Lambda_{\rm -a},\avec_{{\rm gw},a}|\tresid_{-a})
        \nonumber
        \\
        &= \int \textrm{d}\bm\Lambda_{\rm gw} \textrm{d}\bm \Lambda_{a}  \, \textrm{d}\bm \Lambda_{-a}  \, \textrm{d}\avec_{{\rm gw},a} \,\, p(\tresid_a | \avec_{{\rm gw},a},\bm\Lambda_{a}) p(\bm\Lambda_{a})p(\avec_{{\rm gw},a}|\bm\Lambda_{\rm gw},\bm\Lambda_{\rm -a},\tresid_{-a})p(\bm\Lambda_{\rm gw},\bm\Lambda_{\rm -a}|\tresid_{-a})\nonumber
        \\
        &= \int \textrm{d}\bm\Lambda_{\rm gw}\,\textrm{d}\bm\Lambda_{\rm -a} \left[ \int\textrm{d}\bm \Lambda_{a} \,\textrm{d}\avec_{{\rm gw},a} \,\, p(\tresid_a | \avec_{{\rm gw},a}\bm\Lambda_{a}) p(\bm\Lambda_{a})p(\avec_{{\rm gw},a}|\bm\Lambda_{\rm gw},\bm\Lambda_{\rm -a},\tresid_{-a})\right]p(\bm\Lambda_{\rm gw},\bm\Lambda_{\rm -a}|\tresid_{-a})\,,
        \label{eq:pta_loo_posterior_predictive_general}
    \end{align}
\end{widetext}
where the ``HD" subscript signifies that we have assumed Hellings--Downs correlations.
In the second line we have used the definition of conditional probabilities $p(\bm\Lambda_{a},\bm\epsilon_{a},\avec_{a})=p(\bm\Lambda_{a})p(\bm\epsilon_{a},\avec_{a}|\bm\Lambda_{a})$ and in the third line we have marginalized over $\bm\epsilon_{a},\avec_{a}$ following Eq.~\eqref{eq:hyper_parameter_posterior}. 
In the third line we have again used conditional probabilities $p(\bm\Lambda_{\rm gw},\bm\Lambda_{\rm -a},\avec_{{\rm gw},a}|\tresid_{-a})=p(\avec_{{\rm gw},a}|\bm\Lambda_{\rm gw},\bm\Lambda_{\rm -a},\tresid_{-a})(\bm\Lambda_{\rm gw},\bm\Lambda_{\rm -a}|\tresid_{-a})$ and in the last line we re-organize the integrals.
The first term in the integral, $p(\tresid_a | \avec_{\textrm{gw}, a},\bm\Lambda_{a})$, is given by Eq.~\ref{eq:pre_fit_resids_likelihood_gp_coefficients} after (analytically) marginalizing over the intrinsic noise Gaussian process coefficients, $p(\avec_{\textrm{gw}, a} | \bm\Lambda_{\rm gw},\bm\Lambda_{\rm -a}, \tresid_{-a})$ is a Gaussian with mean and covariance given by Eqs.~\ref{eq:bvec_max_likelihood} and~\ref{eq:bvec_covariance_matrix}, $p(\bm\Lambda_{a})$ is the prior on $\bm\Lambda_{a}$, while $p(\bm\Lambda_{\rm gw},\bm\Lambda_{\rm -a}|\tresid_{-a})$ is the posterior of the hyperparameters. 

A simplified version of Eq.~\eqref{eq:pta_loo_posterior_predictive_general} can be obtained if we ignore the Hellings--Downs correlations and assume that the pulsars are only subject to an uncorrelated common process, denoted as ``CN" in equations below. Then
\begin{widetext}
    \begin{align}
        p_{{\rm CN}}(\tresid_a | \tresid_{-a})
        &= \int \textrm{d}\bm\Lambda_{\rm gw}  \textrm{d}\bm\Lambda_{\rm -a} \left[ \int\textrm{d}\bm \Lambda_{a} \,\textrm{d}\avec_{{\rm gw},a} \,\, p(\tresid_a | \avec_{{\rm gw},a}\bm\Lambda_{a}) p(\bm\Lambda_{a})p(\avec_{{\rm gw},a}|\bm\Lambda_{\rm gw}, \bm\Lambda_{\rm -a},\tresid_{-a})\right]p(\bm\Lambda_{\rm gw}, \bm\Lambda_{\rm -a}|\tresid_{-a})\nonumber 
        \\
        &= \int \textrm{d}\bm\Lambda_{\rm gw} \textrm{d}\bm\Lambda_{\rm -a}\left[ \int\textrm{d}\bm \Lambda_{a} \,\textrm{d}\avec_{{\rm gw},a} \,\, p(\tresid_a | \avec_{{\rm gw},a}\bm\Lambda_{a}) p(\bm\Lambda_{a})p(\avec_{{\rm gw},a}|\bm\Lambda_{\rm gw})\right]p(\bm\Lambda_{\rm gw},\bm\Lambda_{\rm -a}|\tresid_{-a}) \nonumber 
        \\
        &= \int \textrm{d}\bm\Lambda_{\rm gw} \left[ \int\textrm{d}\bm \Lambda_{a} \,\, p(\tresid_a | \bm\Lambda_{\rm gw},\bm\Lambda_{a}) p(\bm\Lambda_{a})\right]p(\bm\Lambda_{\rm gw}|\tresid_{-a}) 
        \,,
        \label{eq:pta_loo_posterior_predictive_crn}
    \end{align}
\end{widetext}
where in the second line we have simplified $p(\avec_{{\rm gw},a}|\bm\Lambda_{\rm gw},\bm\Lambda_{\rm -a},\tresid_{-a})=p(\avec_{{\rm gw},a}|\bm\Lambda_{\rm gw})$ due to the lack of Hellings--Downs correlations, and in the third line we have marginalized over $\avec_{{\rm gw},a},\bm\Lambda_{\rm -a}$ following Eq.~\eqref{eq:hyper_parameter_posterior}.

Equations~\eqref{eq:pta_loo_posterior_predictive_general} and~\eqref{eq:pta_loo_posterior_predictive_crn} are estimated as follows. The integral over $\textrm{d}\bm\Lambda_{\rm gw}$ (and $\bm\Lambda_{\rm -a}$ if applicable) is performed through Monte-Carlo integration using $N_s$ samples
\begin{align}
    \bm\Lambda^{s}_{\rm gw},\bm\Lambda^s_{\rm -a} &\sim p(\bm\Lambda_{\rm gw},\bm\Lambda_{\rm -a} | \tresid_{-a})\,,\label{eq:LambdagwMonteCarlo}
\end{align}
from the analysis of Sec.~\ref{sec:datasets}:
\begin{widetext}
\begin{align}
   p_{\textrm{HD}}(\tresid_a | \tresid_{-a}) 
    & \approx \frac{1}{N_s}\sum_{s} \int \textrm{d}\bm\Lambda_{a}\textrm{d}\avec_{{\rm gw},a} p(\tresid_a | \bm\Lambda_{a},\avec_{\textrm{gw}, a}^{s}) p(\avec_{\textrm{gw}, a}^{s} | \bm\Lambda_{\rm gw}^{s},\bm\Lambda^s_{\rm -a}, \tresid_{-a})p(\bm\Lambda_{a})\,, \label{eq:pta_loo_posterior_hd_monte_carlo}
    \\
    p_{\textrm{CN}}(\tresid_a | \tresid_{-a}) &\approx \frac{1}{N_s}\sum_{s}\int \textrm{d}\bm\Lambda_{a}p(\tresid_a | \bm\Lambda_{a},\bm\Lambda_{\rm gw}^{s})p(\bm\Lambda_{a})\,.
    \label{eq:crn_monte_carlo_ppl}
\end{align}
\end{widetext}
The integral over $\textrm{d}\bm \Lambda_{a}$ is performed numerically. The integral over $\textrm{d}\avec_{{\rm gw},a}$ is performed analytically as both terms involving $\avec_{{\rm gw},a}$ are Gaussians.

The above equations require estimating $N_{\rm p}$ posteriors $p(\bm\Lambda | \tresid_{-a})$ -- one for each individual pulsar, $a$. This results in a heavy computational cost that may be unfeasible. 
Instead, if the hyperparameter posterior is not strongly affected by any individual pulsars, we can approximate Eq.~\eqref{eq:LambdagwMonteCarlo} with
\begin{align}
    p(\bm\Lambda | \tresid_{-a}) &= p(\bm\Lambda_{\rm gw},\bm\Lambda_{-a} | \tresid_{-a})p(\bm\Lambda_a)\nonumber\\
    & \approx p(\bm\Lambda_{\rm gw},\bm\Lambda_{-a} | \tresid)p(\bm\Lambda_a)\,.
\end{align}
Crucially, while we use the data from pulsar $a$ to constrain $\bm\Lambda_{\rm gw}$ by assuming that the effect is small, we do not use the same data to constrain $\bm\Lambda_a$, instead still integrating over the prior. We have checked that this approximation has a minor impact on our results while greatly reducing computational cost,  so we adopted it to produce the results in Secs.~\ref{sec:results-parameters} and \ref{sec:results_timing_residuals}.

\bibliography{main}

\end{document}